\documentclass[aps,prb,twocolumn,notitlepage,floatfix]{revtex4-1}
\usepackage{amsmath,amssymb,gensymb,url,hyperref,bbold,color}
\usepackage{mathptmx}
\usepackage{graphicx}
\usepackage{hyperref}
\DeclareGraphicsExtensions{.pdf,.png}
\def\nicefrac#1#2{\genfrac{}{}{}{1}{#1}{#2}}
\def\text#1{\mbox{\scriptsize #1}}
\def\ket#1{\mbox{$\displaystyle\vert\,#1\,\rangle$}}
\def\bra#1{\mbox{$\displaystyle\langle\,#1\,\vert$}}

\def\dyadic#1{\overset{\text{\scriptsize$\leftrightarrow$}}{#1}}
\def\der#1#2{\mbox{$\displaystyle\frac{d #1}{d #2}$}}
\def\vv{\mathbf{v}}
\def\vr{\mathbf{r}}
\def\vp{\mathbf{p}}
\def\vk{\mathbf{k}}
\def\vz{\mathbf{z}}

\def\vl{\mathbf{l}}
\def\vj{\mathbf{j}}
\def\vk{\mathbf{k}}

\def\ve{\mathbf{e}}
\def\vF{\mathbf{F}}
\def\vP{\mathbf{P}}
\def\vE{\mathbf{E}}

\def\He{$^{3}$He}
\def\Hea{$^{3}$He-A}
\def\Heb{$^{3}$He-B}
\def\cE{\mathcal{E}}
\def\cG{\mathcal{G}}

\def\vB{\mathbf{B}}
\def\vL{\mathbf{L}}

\def\TN{T_{\text{N}}}
\def\GN{G_{\text{N}}}
\def\GR{\mathfrak{G}^{\text{R}}}
\def\FR{\mathfrak{F}^{\text{R}}}
\def\be{\begin{equation}}
\def\ee{\end{equation}}
\def\ber{\begin{eqnarray}}
\def\eer{\end{eqnarray}}
\def\ns{\negthickspace}
\definecolor{red}{rgb}{1,0,0}

\begin{document}
\title{
\vspace*{-2cm}
{\footnotesize
\hspace*{-4.0cm} Published in {\it Phys. Rev. B 96, 064511 (2016)}
                 [\href{http://dx.doi.org/10.1103/PhysRevB.94.064511}{doi:10.1103/PhysRevB.94.064511}]
}
\\
\vspace*{2cm}
Electron bubbles and Weyl Fermions in chiral superfluid \Hea
}
\author{Oleksii Shevtsov}
\email{oleksii.shevtsov@northwestern.edu}
\author{J.~A. Sauls}
\email{sauls@northwestern.edu}
\affiliation{Department of Physics and Astronomy, Northwestern University, Evanston, IL 60208 USA}
\date{\today}
\begin{abstract}
Electrons embedded in liquid \He\ form mesoscopic bubbles with radii large compared to the interatomic 
distance between \He\ atoms, voids of $N_{\text{bubble}}\approx 200$ \He\ atoms, generating a negative 
ion with a large effective mass that scatters thermal excitations.
Electron bubbles in chiral superfluid \Hea\ also provide a local probe of the ground state.
We develop scattering theory of Bogoliubov quasiparticles by negative ions embedded in \Hea\ that 
incorporates the broken symmetries of \Hea, particularly broken symmetries under time-reversal 
and mirror symmetry in a plane containing the chiral axis $\hat\vl$.
Multiple scattering by the ion potential, combined with branch conversion scattering by the chiral 
order parameter, leads to a spectrum of Weyl Fermions bound to the ion that support a mass 
current circulating the electron bubble - the mesoscopic realization of chiral edge currents 
in superfluid \Hea\ films.
A consequence is that electron bubbles embedded in \Hea\ acquire angular momentum, 
$\vL\approx -(N_{\text{bubble}}/2)\hbar\,\hat\vl$, inherited from the chiral ground state.
%
%
We extend the scattering theory to
calculate the forces on a moving electron bubble, both the Stokes drag and a transverse force,
$\vF_{\text{W}} = \frac{e}{c}\vv\times\vB_{\text{W}}$, defined by an effective magnetic 
field, $\vB_{\text{W}}\propto\hat\vl$, generated by the scattering of thermal quasiparticles off the 
spectrum of Weyl Fermions bound to the moving ion. The transverse force is responsible for the 
anomalous Hall effect for electron bubbles driven by an electric field reported by the RIKEN 
group. Our results for the scattering cross section, drag and transverse forces on moving 
ions are compared with experiments, and shown to provide a quantitative understanding of the
temperature dependence of the mobility and anomalous Hall angle for electron bubbles in
normal and superfluid \Hea. We also discuss our results in relation to earlier work 
on the theory of negative ions in superfluid \He.
\end{abstract}
\maketitle
\section{Introduction}

A unique feature of the chiral phase of superfluid \He, predicted early on by Anderson and Morel (AM), is 
that this fluid should possess a macroscopic ground-state angular momentum, 
$\vL = L_z\,\hat\vl$,\cite{and60,vol75,cro77,ish77,leg78,mcc79} 
where $\hat\vl$ is the chiral axis along which the Cooper pairs have angular momentum $\hbar$. 
Ground state currents and angular momentum are signatures of broken time-reversal and parity
(BTRP) derived from the orbital motion of the Cooper pairs in \Hea.
In this article we discuss signatures of BTRP generated by the structure of electrons embedded in 
superfluid \Hea. An electron forms a void, a ``bubble'', in liquid \Hea\ that disturbs the chiral 
ground state.\cite{rai77} We show that multiple scattering of Bogoliubov quasiparticles off the electron 
bubble leads to the formation of chiral Fermions bound to the electron bubble, and to a ground state 
angular momentum and mass current circulating each electron bubble. Indeed the electron bubble provides a 
mesocopic realization of chiral edge currents in superfluid \Hea.
A main result of the work reported here is our formulation of transport theory for negative
ions that correctly accounts for the chiral symmetry of superfluid \Hea.
This allows us to show that the chiral structure of the electron bubble in \Hea\ provides a quantitative
theory for the anomalous Hall effect reported by Ikegami et al. \cite{ike13,ike15}
We start with a brief introduction intended to make the connection between ground state currents, angular 
momentum, and chiral edge states in \Hea, with the structure of electron bubbles in \Hea. 

%
The angular momentum of bulk \Hea\ has so far not been measured, perhaps in part because of 
variations in the literature on the magnitute of $L_z$ (c.f. Ref.~[\onlinecite{sau11}] and references therein).
McClure and Takagi (MT) obtained the result, $L_z = (N/2)\,\hbar$, for $N$ atoms confined in a container
of volume $V$ with cylindrical symmetry, and condensed into a chiral p-wave bound state of Fermion pairs. 
This result is independent of whether or not the ground state is a condensate of overlapping chiral 
Cooper pairs ($\xi \gg a$) or a Bose-Einstein condensate of tightly bound chiral molecules 
($\xi\ll a $), where $\xi$ represents the radial extent of the pair wave function, $a=1/\sqrt[3]{n}$ 
is the interatomic spacing and $n$ is the mean atomic density.
While the MT result is in accord with expectations for a BEC of $N/2$ chiral molecules, 
the MT result for the BCS limit suggests that the currents responsible for a ground-state angular momentum 
of $(N/2) \hbar$ are confined on the boundary walls.\cite{vol81,sau11}

%
The existence of currents confined on the boundary is a natural conclusion of the bulk-boundary correspondence 
for a topological phase with broken time-reversal symmetry.\cite{hat93,vol16,miz16} 
In the quasi-2D limit the chiral A-phase is fully gapped and belongs to a topological class 
related to that of integer quantum Hall systems.\cite{rea00,vol88,vol92}
The topology of the chiral AM state requires gapless Weyl fermions confined on the edge of a 
thin film of superfluid \Hea.\cite{vol92,volovik92} For an isolated boundary a branch of Weyl fermions disperses 
linearly with momentum $p_{||}$ along the boundary, i.e. $\varepsilon(p_{||}) = c\,p_{||}$, where 
$c=v_f\,|\Delta|/E_f\ll v_f$ is the velocity of the Weyl Fermions. 
The asymmetry of the Weyl branch under time-reversal, $\varepsilon(-p_{||})=-\varepsilon(p_{||})$, implies 
the existence of a ground-state edge current derived from the occupation of the negative energy 
states.\cite{sto04,sau11} 
%
%
%
For \Hea\ confined in a thin cylindrical cavity, or a film, the \emph{chiral edge current} on the outer
boundary edge, $J=\nicefrac{1}{4}\,n\,\hbar$,\footnote{This is a sheet current obtained by integrating 
the current density confined on the boundary.} is the source of the ground-state angular momentum, 
$L_z = (N/2)\,\hbar$, predicted by MT.\cite{sau11,tsu12,sto04}

\begin{figure}[t]
\centering
\includegraphics[width=0.8\columnwidth,keepaspectratio]{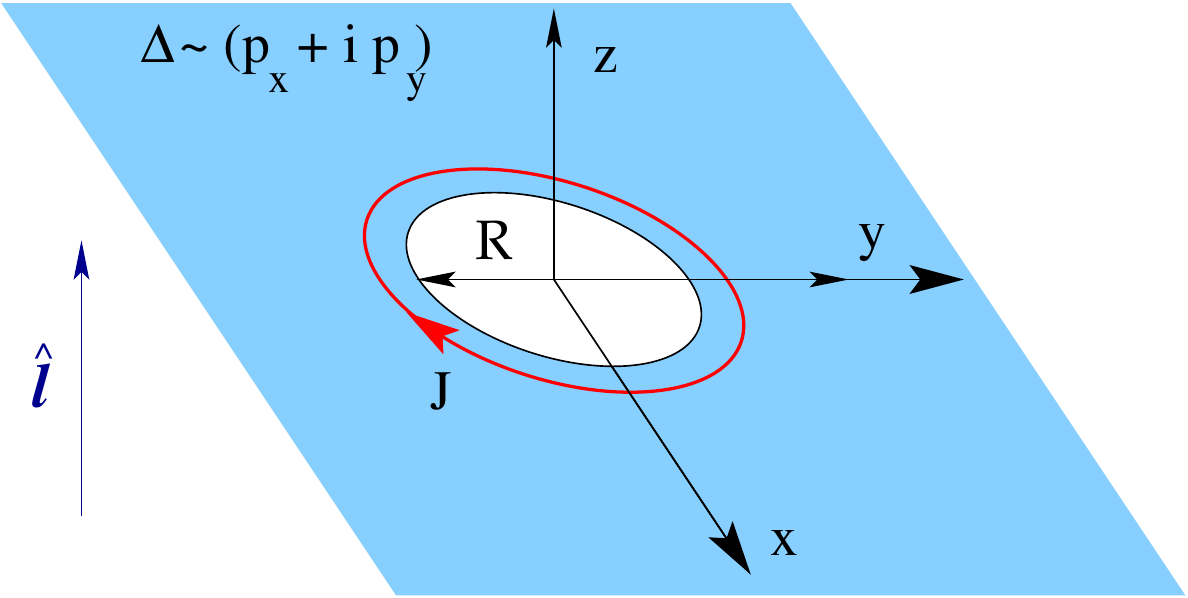}
\caption{
(Color online) An unbounded \Hea\ film with an inner boundary of radius $R\gg \xi_0$. A ground-state edge 
current of magnitude $J=\nicefrac{1}{4}\,n\,\hbar$ circulates on the inner boundary generating an angular momentum 
$L_z = - (N_{\text{hole}}/2)\,\hbar$ with $N_{\text{hole}}= n\,\pi\,R^2\,w$, i.e. the number of \He\ atoms 
\emph{excluded} by the hole.
}
\label{figure-film_hole}
\end{figure}

To reveal the edge currents, consider an
unbounded thin film of superfluid \Hea\ with a circular barrier, a ``hole'', excluding \He\ as 
shown in Fig.~(\ref{figure-film_hole}). The edge current is confined to the boundary on the 
scale of $\xi_{\mbox{\small $\Delta$}}=\hbar v_f/2\Delta\approx 100\,\mbox{nm}$. 
The angular momentum resulting from the edge current circulating the 
hole is, $L_z = -(N_{\text{hole}}/2)\,\hbar$, which is opposite to the chirality of the ground state Cooper 
pairs, and with magnitude given by $N_{\text{hole}}=n\,\pi\,R^2\,w$, the number of \He\ atoms \emph{excluded} 
by the hole of radius $R$ and thickness $w$. 
Nature provides us with such a ``hole'' in the form of an electron bubble to reveal the BTRP of \Hea, and 
to probe the spectrum of chiral edge states, the mass current circulating the electron bubble, and the effect 
of the chiral edge states on the tranport properties of the electron bubble in \Hea.

We begin with the structure of the electron bubble in the normal Fermi liquid phase of \He\ 
in Sec.~(\ref{sec-Ion_Structure-3He-Normal}). The normal-state $t$-matrix and 
scattering phase shifts for quasiparticles scattering off the electron bubble are central to
understanding the properties of the electron bubble in superfluid \Hea.
In Sec.~(\ref{sec-Ion_Structure-3He-A}) we develop scattering theory to calculate the spectrum of chiral Fermions 
bound to the electron bubble in \Hea. We present results for the mass current and orbital angular momentum 
obtained from the Fermionic spectrum.
The momentum and energy resolved differential cross section for the scattering of Bogoliubov quasiparticles 
is developed in Sec.~(\ref{sec-Scattering_Theory}), and used to calculate the forces on electron bubbles 
moving in the chiral phase of superfluid \He. 
We present new theoretical predictions and analysis for the drag force on electron bubbles in \Hea, and 
particularly the transverse force responsible for the anomalous Hall current of electron bubbles in 
superfluid \Hea. 
In Sec.~(\ref{sec-Results_Electron-Mobility}) we present the quantitative comparison of our theory with the 
measurements of the drag force and anomalous Hall effect reported by Ikegami et al. \cite{ike13,ike15} 
Our analysis establishes that the observation of the anomalous Hall effect for negative ions 
is not only a signature of BTRP, but a signature of chiral Fermions circulating 
the electron bubble.

We point out that previous theories for the mobility of ions in superfluid \Hea\ start from an 
implicit assumption of mirror symmetry in the formulation of the transport cross-section for 
scattering of quasiparticles off the electron bubble.
Specifically, in Sec.~(\ref{sec-Microscopic_Reversibility}) and App.~(\ref{appendix_Salmelin-errors}) we discuss 
our theory in relation to the earlier theoretical works of Salomaa et al.\cite{sal80} and Salmelin et 
al.,\cite{sal89,sal90} and point out that these earlier theoretical works give \emph{zero} Hall mobility 
(Ref.~[\onlinecite{sal90}]), or report a spurious Hall 
mobility that is an artefact of an error in evaluating the kinematics for the scattering of quasiparticles 
off the ion. As a result, the theory reported in Refs. [\onlinecite{sal89,sal90}] not only predicts a spurious 
Hall mobility in \Hea, but also a spurious anisotropic mobility in normal liquid \He.

Our formulation of the transport theory correctly accounts for the chiral symmetry of superfluid \Hea, 
which is at the root of the anomalous Hall effect for electrons in \Hea,\footnote{{For a historical review of 
theories of the anomalous Hall effect in solid state systems see N.~A. Sinitsyn, J. Phys. Cond. 
Matt., ~20, ~023201, (2008).}} and as shown in Sec.~(\ref{sec_Experiment-Theory}) is in quantitative agreement 
with the experimental measurements reported in Refs. [\onlinecite{ike13,ike15}].

\section{Electron bubbles in Liquid \He}\label{sec-Ion_Structure-3He-Normal}

Electrons experience a repulsive barrier $\approx 1\,\mbox{eV}$ at the surface of 
liquid Helium.\cite{woo65} When an electric field pushes the electron into Helium the combination of the 
barrier, the surface tension and zero-point kinetic energy of the electron conspire to form 
a self-trapped electron in a spherical void of radius $R$, an ``electron 
bubble''.\cite{fer57,kup61,fet76,dobbs00} 
The basic model of an electron bubble in liquid \He\ is based on an energy function that 
consists of three terms,\cite{fer57,aho78}
\begin{align}\label{E_bubble}
E(R,P) = E_0(U_0,R) + 4\pi R^2\gamma + \frac{4\pi}{3}R^3P,
\end{align}
where $\gamma = 0.15\,\mathrm{erg/cm^2}$ is the surface tension of \He,\cite{lov55,suz88} 
$P$ is the external pressure and $E_0$ is the ground state energy of the electron bubble trapped 
in an isotropic potential of radius $R$ and depth $-U_0$.
In the limit $U_0\rightarrow\infty$, $E_0=-U_0+\pi^2\hbar^2/2m_e\,R^2$ is the energy of the electron
of mass $m_e$ in its ground state. The balance between the surface tension of liquid \He, the external 
pressure and the kinetic energy of the confined electron determines the bubble radius,
$P = {\pi\hbar^2}/{4 m_e R^5} - {2\gamma}/{R}$.
For zero pressure the radius is then
\be
R=\left(\frac{\pi\hbar^2}{8 m_e\gamma}\right)^{\nicefrac{1}{4}} \approx 2.38\,\mbox{nm}
\,.
\ee
The bubble radius is large compared to the Fermi wavelength of \He\ quasiparticles, 
$\lambda_f=1/k_f=0.127\,\mbox{nm}$, set by the \He\ density, but is small compared to the Cooper pair 
correlation length, $\xi_0=\hbar v_f/2\pi\,k_{B}T_c\approx 77.3\,\mbox{nm}$.
It is useful to refer to the dimensionless ratio, $k_fR$, which for the infinte barrier limit 
is $k_fR =18.67$ at $P=0$.
Models for the confining potential with a finite pressure-dependent $U_0\sim 1\,\mbox{eV}$ 
yield a slightly smaller radius of $k_fR\approx 16.74$.\cite{aho78}

\subsection{Electron Mobility in Normal \He}

A different measure of the size of the electron bubble may be obtained from the scattering 
of \He\ quasiparticles off the electron bubble, i.e. the total cross section presented to 
quasiparticles with momenta and energies near the Fermi surface. The scattering of quasiparticles 
off the heavy electron bubble determines the mobility of the electron bubble.
The heavy mass of the ion and large cross section for quasiparticle collisions
imply that the scattering of quasiparticles off the electron bubble is nearly 
elastic.\cite{jos69,fet77} The mobility of the electron bubble 
in normal \He\ is then temperature independent over the range 
$T_c<T<T_0=\hbar^2k_f^2/2M\approx 30\,\mbox{mK}$, and given by 
\be\label{eq-mobility_normal-state}
\frac{e}{\mu_{\text{N}}} = n_3 p_f\sigma_{\text{N}}^{\text{tr}}
\,,\quad
\sigma_{\text{N}}^{\text{tr}} = 
\int d\Omega_{\hat\vk'}\frac{d\sigma}{d\Omega_{\hat\vk'}} \left(1-\hat\vk'\cdot\hat\vk\right)
\,,
\ee
where $\sigma_{\text{N}}^{\text{tr}}$ is the transport cross section for elastic scattering of 
quasiparticles off an electron bubble, and 
\be\label{eq-dsigma_dOmega}
\frac{d\sigma}{d\Omega_{\hat\vk'}} = 
	\left\vert\frac{m^*}{2\pi\hbar^2}\,t^{\text{R}}_{\text{N}}(\hat\vk',\hat\vk;E)\right\vert^2
\,,
\ee
is the differential cross-section defined by the on-shell $t$-matrix for normal-state 
quasiparticles with effective mass $m^*$ scattering off a static electron bubble.

The full $t$-matrix obeys the Lippmann-Schwinger equation, 
$\TN^{\text{R}}=V+VG_{\text{N}}^{\text{R}}\TN^{\text{R}}$, 
where $G^{\text{R}}_{\text{N}}$ is the retarded propagator for Fermions in the normal Fermi liquid.
At temperatures $k_{\text{B}}T\ll E_f$ the properties of \He\ are dominated by quasiparticles with
momenta near the Fermi surface, $\vk\simeq k_f\hat\vk$, and excitation energies, 
$\xi_{\vk}\simeq v_f(|\vk|-k_f)$ with $|\xi_{\vk}| \ll E_f$. The corresponding $t$-matrix
describing the scattering of quasiparticles off the electron bubble is obtained by 
separating the propagator as $\GN^{\text{R}} = \GN^{\text{R,low}} + \GN^{\text{R,high}}$, where 
$\GN^{\text{R,low}}=a/(E+i0^{+} - \xi_{\vk})$ is the low-energy quasiparticle propagator with
residue $a$, and $\GN^{\text{R,high}}$ is the high-energy, incoherent propagator. The latter 
renormalizes the bare \He-Ion interaction, $U = V + V\GN^{\text{R,high}}U$. The resulting $t$-matrix,
$\bra{\vk'}\TN^{\text{R}}\ket{\vk}\equiv t^{\text{R}}_{\text{N}}(\hat\vk',\hat\vk;E)$, for elastic scattering
of low-energy quasiparticles with energy $E$, and momenta $\vk=k_f\hat{\vk}$ to $\vk'=k_f\hat{\vk}'$
on the Fermi surface is then
\ber
\hspace*{-2mm}
t_{\text{N}}^{\text{R}}(\hat\vk',\hat\vk;E) &=& u(\hat\vk',\hat\vk) 
\nonumber\\
&+&
N_f\ns
\int\frac{d\Omega_{\vk''}}{4\pi}\ns
u(\hat\vk',\hat\vk'') 
g^{\text{R}}_{\text{N}}(\hat\vk'',E)\,
t^{\text{R}}_{\text{N}}(\hat\vk'',\hat\vk;E)
\,,
\eer
where $N_f=m^*k_f/2\pi^2\hbar^2$ is the single-spin density of states at the Fermi surface,
$m^*=p_f/v_f$ is the quasiparticle effective mass, 
$g^{\text{R}}_{\text{N}}(\hat\vk,E)=\nicefrac{1}{a}\int\,d\xi_{\vk}\,\GN^{\text{R,low}}(\vk,E)=-i\pi$
is the quasiclassical progragator, and $u(\vk',\vk) = \bra{\vk'}U\ket{\vk}$. For a spherically symmetric 
electron bubble the quasiparticle-ion interaction and the $t$-matrix can be expanded 
as, $u(\hat\vk',\hat\vk)=\sum_{l\ge 0}(2l+1)\,u_l\,P_l(\hat\vk'\cdot\hat\vk)$, 
and similarly for $t^{\text{R}}_{\text{N}}(\hat\vk',\hat\vk)$, where $\{P_l(x)\,|\,l=0,1,2,\ldots\}$ 
is the complete set of Legendre polynomials. Using the convolution integral,
$\int\frac{d\Omega_{\vk''}}{4\pi}\,P_{l'}(\hat\vk'\cdot\hat\vk'')\,P_{l}(\hat\vk''\cdot\hat\vk)
 =\delta_{ll'}\,P_l(\hat\vk'\cdot\hat\vk)/(2l+1)$, we obtain $t^{\text{R}}_{l}(E) = u_l/(1+i\pi N_f\,u_l)$.
The structure of the $t$-matrix can be encoded in the scattering phases shifts, $\delta_l$, 
defined in terms of the strength of the quasiparticle-ion potential in each angular momentum channel, 
$u_l$, and the density of states, $N_f$; $\tan\delta_l = -\pi N_f\,u_l$, with the $t$-matrix 
expressed as,
\be\label{t_N}
t^{\text{R}}_{\text{N}}(\hat\vk',\hat\vk;E) = 
-\frac{1}{\pi N_f}\sum_{l=0}^{\infty}(2l+1)e^{i\delta_l}\sin\delta_l\,P_l(\hat{\vk}'\cdot\hat{\vk})
\,.
\ee
Integrating Eq.~(\ref{eq-dsigma_dOmega}) over all scattering directions, we obtain the standard result 
for the total cross section\cite{messiah58}
\ber\label{sigma_N0}
\sigma_{\text{N}}
= \frac{4\pi}{k_f^2}
\sum_{l=0}^{\infty}\left[(2l+1)\sin^2\delta_l \right]
\,.
\eer
Similarly, the transport cross section is determined by the set of scattering phase shifts that
parametrize the quasiparticle-ion potential,
\ber\label{sigma_N}
\sigma_{\text{N}}^{\text{tr}} 
= \frac{4\pi}{k_f^2}
\sum_{l=0}^{\infty}
&&\left[(2l+1)\sin^2\delta_l \right.
\nonumber\\
&&\left. 
- 2(l+1)\cos(\delta_{l+1} 
- \delta_l)\sin\delta_{l+1}\sin\delta_l
             \right]
\,.
\eer

\subsection{Hard-Sphere Scattering of Quasiparticles}

\begin{figure}[t]
\centering
\includegraphics[width=0.99\columnwidth,keepaspectratio]{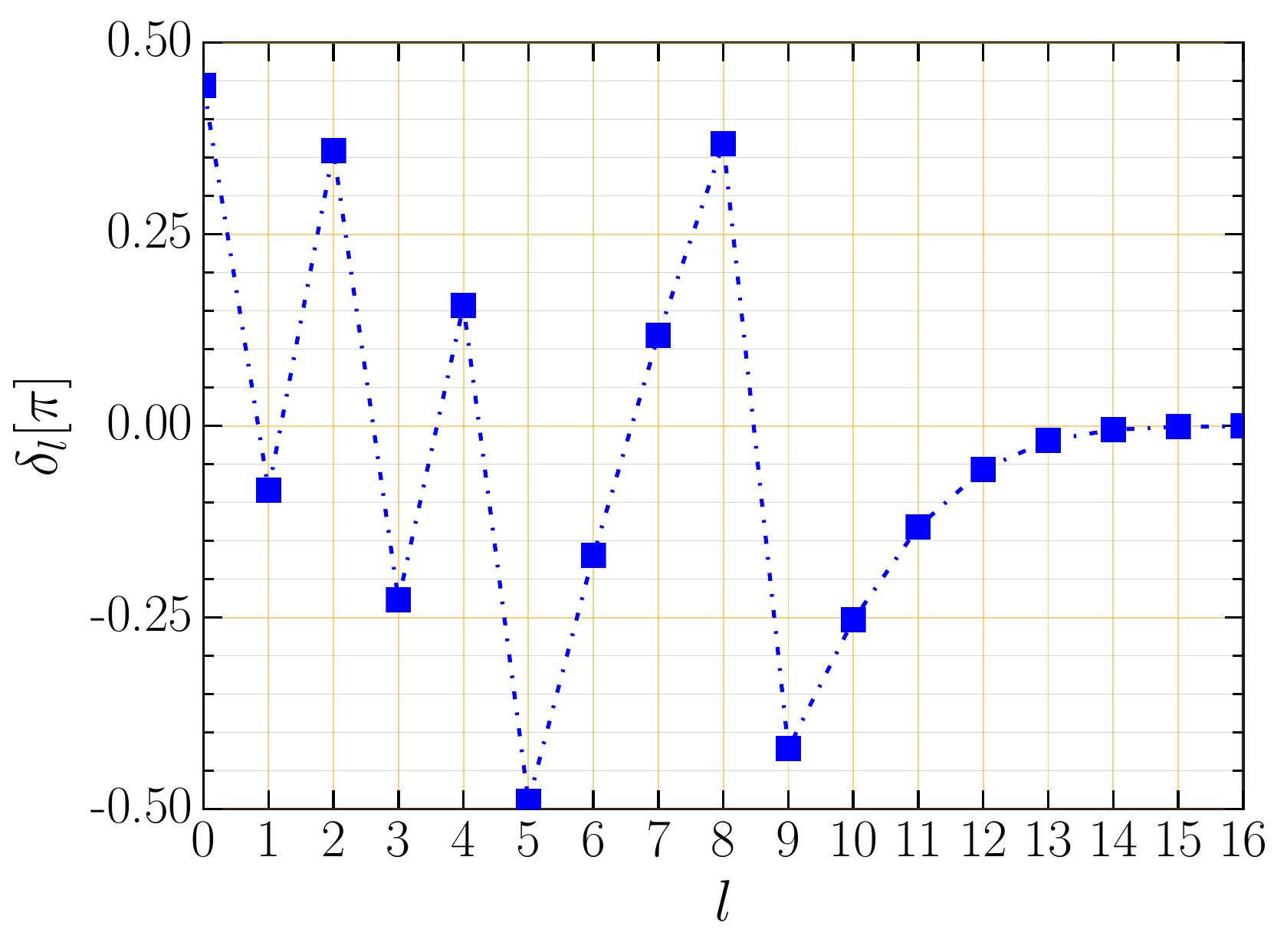}
\caption{Phase shifts as a function of angular momentum channel number $l$ for the hard-sphere 
         potential with $k_fR = 11.17$. Note that channels with $l>12$ are effectively irrelevant.}
\label{figure_hard-sphere_phase-shifts}
\end{figure}

The structure of the electron bubble as a spherical void of displaced \He\ suggests the model 
of a short-range repulsive barrier preventing penetration of \He\ into the bubble. The potential 
barrier, $V_0\approx 1\,\mbox{eV}$, is very large compared to typical quasiparticle kinetic energies,
suggesting a reasonable model for the quasiparticle-ion potential is a single parameter 
hard-sphere potential parametrized by barrier radius $R$.
The scattering phase shifts that define the quasiparticle-ion $t$-matrix for hard-sphere scattering are 
calculated in standard textbooks,\cite{messiah58}
\be\label{delta_l_hs}
\tan\delta_l = \frac{j_l(k_fR)}{n_l(k_fR)}
\,,
\ee
where $j_l(x)$ and $n_l(x)$ are order $l$ spherical Bessel functions of the first and second kind, 
respectively. 

Figure (\ref{figure_hard-sphere_phase-shifts}) shows the set of phase shifts for a hard sphere 
with a ratio of radius to Fermi wavelength of $k_fR=11.17$. Note that for channels with $l\gtrsim k_fR$, 
the phase shift decreases rapidly to zero.
The radius is determined by requiring the transport cross-section computed for the hard-sphere
potential reproduce the measured normal-state ion mobility according to Eqs.~(\ref{eq-mobility_normal-state}),
(\ref{sigma_N}) and (\ref{delta_l_hs}).
At $P=0$ bar the Fermi wave number, $k_f=7.853\,\mbox{nm}^{-1}$, determines the Fermi momentum, 
$p_f=\hbar\,k_f$, and particle density, $n_3= k_f^3/3\pi^2$.
Combined with the measured normal-state mobility,
$\mu_{\text{N}}^{\text{exp}} = 1.7\times 10^{-6}\,\mathrm{m^2/Vs}$,\cite{ike13}
we obtain $k_f\,R =11.17$, smaller than the bubble radius determined by the
surface tension and zero-point kinetic energy of the electron. 
For scattering of quasiparticles off the electron bubble this is the relevant measure of the size of 
the electron bubble.\footnote{A quantitative physical explanation for the difference in these different 
determinations of the size of the electron bubble has not been presented. One possible source
of the discrepancy is the assumption implied by the analysis based on Eq.~(\ref{E_bubble}) that the surface 
tension, $\gamma$, determined in the hydrostatic limit can be extended to curvatures of order 
$R\simeq 2\,\mbox{nm}$.}
In what follows we develop the theory for the structure of the electron bubble in chiral superfluid  
\Hea\ based on multiple scattering of Bogoliubov quasiparticles off the negative ion.

\section{Structure of an Electron Bubble in \Hea}\label{sec-Ion_Structure-3He-A}

The structure of an electron bubble in \Hea\ is much richer than that in normal \He. 
However, multiple scattering channels of electon bubble are central in determining the spectrum 
of chiral Fermions confined near the electron bubble. Here we develop the theory 
for Bogoliubov quasiparticles scattering off an electron bubble embedded in superfluid \Hea,  
and use the scattering theory to calculate the local spectrum of chiral Fermions bound to the
electron bubble, as well as the mass current and angular momentum circulating the electron bubble. 
Our formulation parallels 
Refs.~[\onlinecite{bay77}], [\onlinecite{sal80}], [\onlinecite{thu81}] and [\onlinecite{sal90}]; however, 
we incorporate broken parity and time-reversal, in addition to broken $U(1)$ and $SO(3)$ 
symmetries, of the ground state of \Hea\ in our formulation of the scattering of quasiparticles 
off electron bubbles. 

Fermionic excitations of superfluid \Hea\ are coherent superpositions of normal-state particles and 
holes described by four-component Bogoliubov-Nambu spinor wavefunctions, 
$\Psi(\vr) = (u_{\uparrow}(\vr),u_{\downarrow}(\vr),v_{\downarrow}(\vr),v_{\uparrow}(\vr))^{\mathrm{T}}$,
that are solutions of Bogoliubov's equations
\vspace*{-3mm}
\ber
\widehat{H}_S\Psi(\mathbf{r}) = E\Psi(\mathbf{r}), 
\quad
\widehat{H}_S = \begin{pmatrix} \hat{H}_N & \hat\Delta(\vp) \\ 
                \hat\Delta^{\dagger}(\vp) & -\hat{H}_N \end{pmatrix}
\,,
\\
\hat{H}_N = (\frac{\vp^2}{2m^{\ast}}-\mu)\,\mathbb{1}
\,, \quad
\hat\Delta({\vp}) = 
\sigma_x  
\Delta (\vp_x + i \vp_y)/p_f
\,,
\label{BdG}
\eer
where $\hat\Delta(\vp)$ is the mean-field pairing potential (order parameter) responsible 
for particle-hole coherence of the Fermionic excitations, and for branch conversion
scattering between particle-like and hole-like Bogoliubov quasiparticles.
Note that $\vp=-i\hbar\nabla$, $\mathbb{1}$ is the unit matrix in spin space,
and $\sigma_x$ is the Pauli matrix describing equal-spin pairing state (ESP) of
Cooper pairs with spin projections $S_x=\pm 1$; equivalently, the Cooper pairs have
zero spin projection along $\hat\vz$. The chiral axis $\hat{\vl}$ for
A-phase Cooper pairs is also along $\hat\vz$. 
Thus,  the $4\times 4$ equation splits into a pair of two-component equations for 
$\Psi_{\uparrow}=(u_{\uparrow},0,0,v_{\uparrow})^{\mathrm{T}}$ and 
$\Psi_{\downarrow}=(0,u_{\downarrow},v_{\downarrow},0)^{\mathrm{T}}$. 

\subsection{Scattering States and Propagators}

The scattering states are Bogoliubov quasiparticles in homogeneous \Hea, i.e. 
far from the electron bubble, in which case the orbital part of the mean-field 
pairing potential can be expressed as 
$\Delta(\hat\vk_x+i\hat\vk_y)=\Delta\sin\theta\,e^{+i\phi}$, where $\theta$
is the polar angle of the relative momentum of the Cooper pairs in momentum space
and the azimuthal angle, $\phi$, is the phase of the 
Cooper pairs in momentum space that winds by $2\pi$ about the chiral axis, $\hat{\vl}$. 
This phase winding plays a central
role in the scattering of quasiparticles off the electron bubble embedded in \Hea. 
The scattering states are eigenstates of momentum, $\vp\ket{\vk}=\hbar\vk\,\ket{\vk}$. 
There are four Bogoliubov quasiparticle states for each
energy - particle-like and hole-like excitations each with two degenerate spin states.
The Bogoliubov-Nambu spinors for the scattering states have the form
\ber\label{BN_spinor1}
|\Psi_{1,\vk\sigma}\rangle 
&=& \begin{pmatrix} u_{\vk}\chi_{\sigma} \\ -v_{\vk}^{\ast}\chi_{\bar{\sigma}}
    \end{pmatrix}\otimes|\vk\rangle
=|\Phi_{1,\vk\sigma}\rangle\otimes|\vk\rangle
\,,
\\
|\Psi_{2,\vk\sigma}\rangle 
&=& \begin{pmatrix} v_{\vk}\chi_{\sigma} \\ -u_{\vk}^{\ast}\chi_{\bar{\sigma}}
    \end{pmatrix}\otimes|\vk\rangle
= |\Phi_{2,\vk\sigma}\rangle\otimes|\vk\rangle,
\label{BN_spinor2}
\eer
where the particle and hole amplitudes are given by
\ber\label{eq-Bogoliubov_amplitudes}
u_{\vk} = \frac{1}{\sqrt{2}}\sqrt{1+\frac{\xi_k}{E_{\vk}}}\,,\quad
v_{\vk} = \frac{1}{\sqrt{2}}\sqrt{1-\frac{\xi_k}{E_{\vk}}}\,e^{i\phi}
\,,
\\
\langle\mathbf{r}|\vk\rangle = e^{i\vk\mathbf{r}}
\,,\;\chi_{\uparrow} = \begin{pmatrix} 1 \\ 0 \end{pmatrix}
\,,\;\chi_{\downarrow} = \begin{pmatrix} 0 \\ 1 \end{pmatrix}
\,,\;\chi_{\bar{\uparrow}} = \chi_{\downarrow}
\,,
\eer
where $E_{\vk}=\sqrt{\xi_{\vk}^2 + |\Delta(\hat\vk)|^2}$ is the excitation energy for 
Bogoliubov quasiparticles.
The spinors, $|\Psi_{1,\vk\sigma}\rangle$, are the particle-like states with $\xi_k > 0$
and group velocity $\nabla_{\vk}E_{\vk}>0$, while $|\Psi_{2,\vk\sigma}\rangle$ are 
the hole-like states with $\xi_k < 0$ and $\nabla_{\vk}E_{\vk}<0$.
Note that the winding number of the Cooper pairs is imprinted as a relative phase between 
the particle- and hole like amplitudes in Eq.~(\ref{eq-Bogoliubov_amplitudes}).

The causal propagator is the retarded Green's function 
of Bogoliubov's equations, $\left(\varepsilon\widehat{1}-\widehat{H}_{\text{S}}\right)\,
\widehat{G}^{\text{R}}_{\text{S}}=\widehat{1}$, with $\varepsilon=E+i\eta$  
($\eta\rightarrow 0^+$), 
which for the bulk excitations in the homogeneous A-phase is given by 
\be\label{eq-GRS_bulk}
\widehat{G}^{\text{R}}_{\text{S}}(\vk,E) = 
\frac{1}{\varepsilon^2 - E_{\vk}^2}
\begin{pmatrix}
(\varepsilon + \xi_k)\mathbb{1} & -\hat\Delta(\hat{\vk}) 
\\
-\hat\Delta^{\dagger}(\hat{\vk}) & (\varepsilon - \xi_k)\mathbb{1}
\end{pmatrix}
\,.
\ee
Note that $\widehat{G}^{\text{R}}_{\text{S}}(\vk,E)$ is restricted to the
low energy region of the Fermi surface where the normal-state is well described 
by long-lived quasiparticles. The corresponding Nambu matrix for the normal-state 
propagator,
\be\label{eq-GRN_bulk}
\widehat{G}^{\text{R}}_{\text{N}}(\vk,E) = 
\begin{pmatrix}
(\varepsilon - \xi_k)^{-1}\mathbb{1} & 0
\\
0 & (\varepsilon + \xi_k)^{-1}\mathbb{1}
\end{pmatrix}
\,,
\ee
includes both the particle- and hole propagators.

\subsection{T matrix }

The electron bubble introduces a strong, short-range potential that scatters Bogoliubov 
quasiparticles. The $t$-matrix is given by the Lippmann-Schwinger equation, which  
becomes a $4\times 4$ Nambu matrix whose elements define the transition amplitudes for 
scattering of Bogoliubov particles and holes, including branch conversion, i.e. Andreev 
scatteirng,
\be\label{eq-Lippmann-Schwinger}
\widehat{T}_{\text{S}}=\widehat{V}+\widehat{V}\widehat{\cG}^{\text{R}}_{\text{S}}\widehat{T}_{\text{S}}
\,, 
\quad \mbox{where}\quad
\widehat{V}=\begin{pmatrix} V(\vr) & 0 \\ 0 & -V(\vr) \end{pmatrix}
\,,
\ee
is the Nambu matrix for the ion potential,
and $\widehat\cG^{\text{R}}_{\text{S}}$ is the exact propagator in the presence of the local potential 
of the ion. For an ion with small cross-section on the scale of the size of Cooper pairs, we 
are justified in replacing $\widehat\cG^{\text{R}}_{\text{S}}\rightarrow \widehat{G}^{\text{R}}_{\text{S}}$,
i.e. the bulk propagator in the absence of the ion given by Eq.~(\ref{eq-GRS_bulk}).
We can use the corresponding Lippman-Schwinger equation for scattering of quasiparticles in the normal state 
to eliminate the ion potential $\hat{V}$ in favor of the normal-state $t$-matrix,\cite{ser83}
\be\label{eq-Lippmann-Schwinger-Nambu}
\widehat{T}_S = \widehat{T}_N + \widehat{T}_N\left(\widehat{G}_S^R - \widehat{G}_N^R\right)\widehat{T}_S
\,.
\ee
The normal state $t$-matrix can be expressed in terms of the 
quasiparticle $t$-matrix, and has the diagonal form in Nambu space,
\be\label{T_N}
\widehat{T}_{\text{N}}(\hat{\vk}',\hat{\vk}) = 
\begin{pmatrix}
t^{\text{R}}_{\text{N}}(\hat{\vk}',\hat{\vk})\mathbb{1} & 0 
\\
0 & -\left[t^{\text{R}}_{\text{N}}(-\hat{\vk}',-\hat{\vk})\mathbb{1}\right]^{\dagger}
\end{pmatrix}
\,.
\ee

The ground state of \Hea\ breaks rotational symmetry, but preserves axial rotations combined with 
a compensating gauge transformation. Thus, scattering of quasiparticles 
off the electron bubble in \Hea\ no longer separates into angular momentum channels with a precise
$l$. However, the projection of the angular mometum, labelled by $m$, is conserved for non-branch 
conversion scattering, and changes by one unit of angular momentum for branch conversion scattering.
Thus, in reducing the $t$-matrix for the scattering of Bogoliubov quasiparticles in \Hea,
it is convenient to rewrite Eq.~(\ref{t_N}) as an expansion in azimuthal harmonics by 
using the addition theorem to express the Legendre functions in terms of the spherical 
harmonics.\cite{mathews65} We then change the order of the summations over $l$ and $m$,
\ber\label{t_N_2}
t_N(\hat{\vk}',\hat{\vk}) 
&=& 
-\frac{1}{\pi N_F}\sum_{m=-\infty}^{\infty}t_N^m(u',u)e^{-im(\phi'-\phi)},
\\
t_N^m(u',u) 
&=&
4\pi
\sum_{l = |m|}^{\infty}e^{i\delta_l}\sin\delta_l\,
\Theta_l^{m}(u')\Theta_l^{m}(u),
\nonumber
\eer
where $(\theta,\phi)$ [$(\theta',\phi')$] are spherical coordinates of $\hat{\vk}$ [$\hat{\vk}'$] in 
momentum space, with $u \equiv \cos\theta$ and $u' \equiv \cos\theta'$.
The functions $\Theta_{l}^{m}(\cos\theta)$ are 
spherical harmonics with the phase winding removed, i.e. 
$Y_l^m(\theta,\phi) = \Theta_{l}^{m}(\cos\theta)\,e^{im\phi}$.

For elastic scattering of Bogoliubov quasiparticles we can reduce Eq.~(\ref{eq-Lippmann-Schwinger-Nambu})
to a linear integral equation with $\widehat{T}^{\text{R}}_{\text{N}}(\hat{\vk}',\hat{\vk})$ as the source 
term,\footnote{{Eq.~(\ref{eq-Lippmann-Schwinger}) is formulated for all energies and momenta of the 
incident and final state excitations. We require the on-shell $t$-matrix in the low energy region 
near the Fermi surface. High-energy intermediate states are included in the phase shifts defining 
the normal-state $t$-matrix, and can be evaluated for momenta on the Fermi surface and $E=0$.
We note that physical quantities, like the mobility or transport cross section, are 
determined by $\widehat{T}^{\mbox{\scriptsize R}}_S(\hat{\vk}',\hat{\vk})$ with an additional constraint, 
$E\ge|\Delta(\hat{\vk})|$ and $E\ge|\Delta(\hat{\vk}')|$, see Eqs.~(\ref{dsigma-dOmega})-(\ref{sigma_-}). 
Strictly speaking, only these matrix elements are ``on-shell'' since there are no bulk quasiparticle
states with momentum $\vk$ for $E < |\Delta(\hat{\vk})|$. Nevertheless, 
Eq.~(\ref{eq-Lippmann-Schwinger-Nambu}) contains both off-shell and on-shell matrix elements. 
After solving this equation we retain only the on-shell matrix elements.
}}
\ber\label{eq-Lippmann-Schwinger-Quasiclassical}
\hspace*{-3mm}\widehat{T}^{\text{R}}_{\text{S}}(\hat{\vk}',\hat{\vk};E) 
= 
\widehat{T}^{\text{R}}_{\text{N}}(\hat{\vk}',\hat{\vk}) 
+ N_f\ns\int\frac{d\Omega_{\vk''}}{4\pi}\times \qquad\qquad\quad
\nonumber\\
\widehat{T}^{\text{R}}_{\text{N}}(\hat{\vk}',\hat{\vk}'') 
[\widehat{g}^{\text{R}}_{\text{S}}(\hat{\vk}'',E)
-
\widehat{g}^{\text{R}}_{\text{N}}(\hat{\vk}'',E)]
\widehat{T}^{\text{R}}_{\text{S}}(\hat{\vk}'',\hat{\vk};E)
\,,
\eer
where the propagators in Eq.~(\ref{eq-Lippmann-Schwinger-Quasiclassical}) are confined to a narrow shell of 
energies and momenta near the Fermi surface and evaluated in the quasiclassical approximation,\footnote{{
$\int d^3k/(2\pi)^3(\ldots)\approx N_f\int d\Omega_{\vk}/4\pi\int d\xi_k(\ldots)$}}
\ber
\widehat{g}^{\text{R}}_{\text{S}}(\hat{\vk}'',E) 
= -\frac{\pi}{\sqrt{|\Delta(\hat{\vk})|^2 - \varepsilon^2}}
\begin{pmatrix}
\varepsilon\mathbb{1} & -\hat\Delta(\hat{\vk}) 
\\
-\hat\Delta^{\dagger}(\hat{\vk}) & \varepsilon\mathbb{1}
\end{pmatrix}
\,.
\eer

\begin{figure}[!]
\centering
\includegraphics[width=0.9\columnwidth,keepaspectratio]{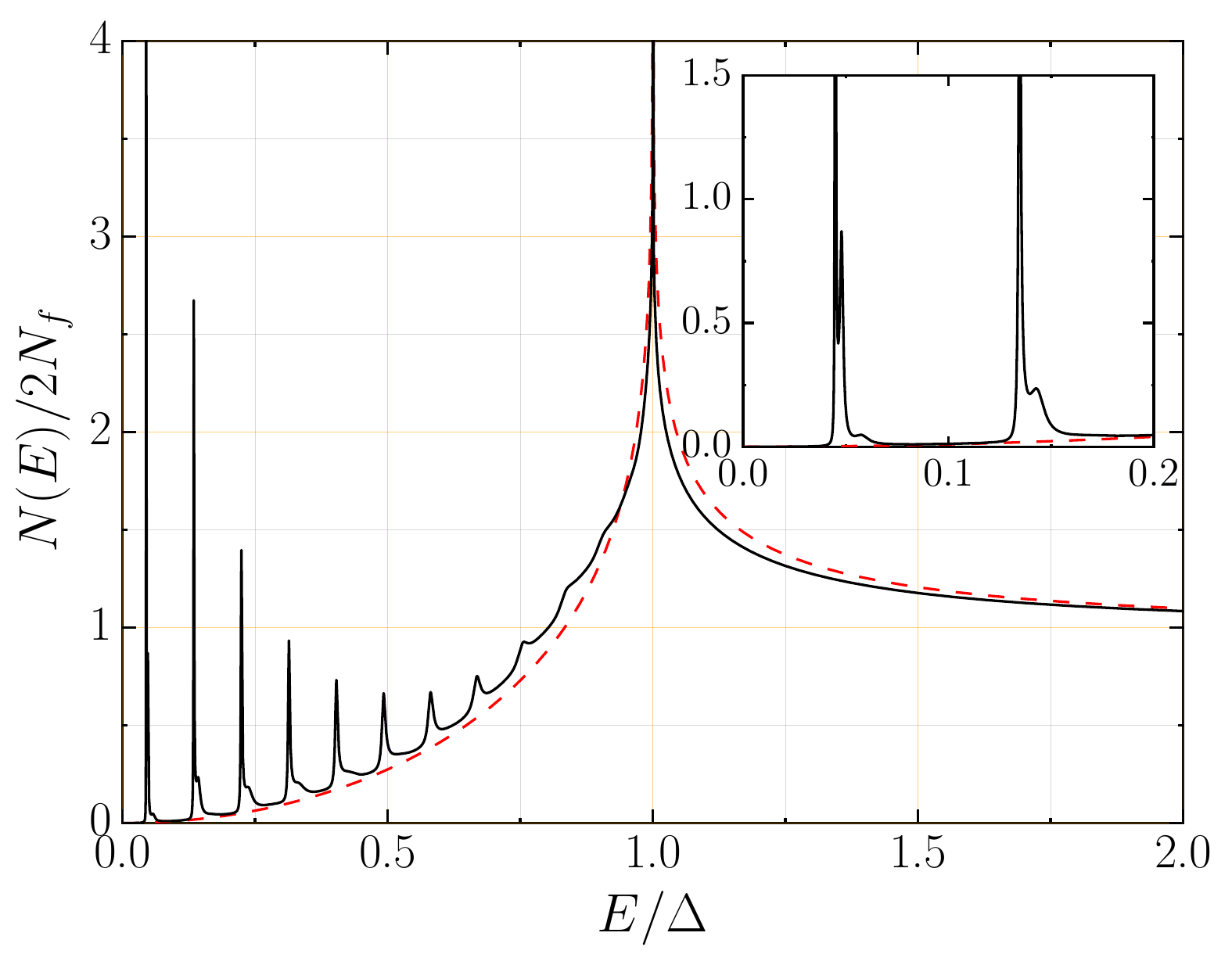}
\caption{
(Color online) Local quasiparticle density of states around an electron bubble in 
superfluid \Hea\ at distance $r = 30\,k_f^{-1}$ from the center of the 
bubble, and in the equitorial plane, i.e. polar angle $\vartheta=\pi/2$ relative 
to the chiral axis. The bubble is 
described as a hard sphere with radius $R = 11.17\,k_f^{-1}$. The inset shows two 
low-energy resonances with internal structure in their spectral density.
}
\label{Fig_LDOS}
\end{figure}

The corresponding normal-state quasiclassical propagator is 
$\widehat{g}^{\text{R}}_{\text{N}}(\vk,E)=-i\pi\widehat{1}$.
For scattering off an electron bubble in \Hea\ the quasiclassical $t$-matrix reduces to a set of $4\times 4$ 
matrix equations for scattering amplitudes, $t^{m}_a$ for Nambu components, $a=1,2,3,4$, for each angular 
momentum quantum number, $m\in\{0\,\pm 1,\pm 2,\ldots\}$. In particular, we parametrize 
$\widehat{T}^{\text{R}}_{\text{S}}$ as
\begin{widetext}
\be
\widehat{T}^{\text{R}}_{\text{S}} 
= \ns
\begin{pmatrix}
t_1(\hat{\vk}',\hat{\vk};E)\mathbb{1}  & t_2(\hat{\vk}',-\hat{\vk};E)\sigma_x 
\cr
t_3(-\hat{\vk}',\hat{\vk;E})\sigma_x   & t_4(-\hat{\vk}',-\hat{\vk};E)\mathbb{1}
\end{pmatrix}
=\ns \frac{-1}{\pi N_F}\sum\limits_{m=-\infty}^{\infty}\ns e^{-im(\phi'-\phi)}\ns
\begin{pmatrix}
t_1^m(u',u)\mathbb{1} & (-1)^me^{i\phi'}t_2^m(u',-u)\sigma_x 
\cr
(-1)^{m+1}e^{-i\phi'}t_3^m(-u',u)\sigma_x & t_4^m(-u',-u)\mathbb{1}
\end{pmatrix}
\,.
\label{T_S}
\ee
The prefactors $(-1)^m$ in Eq.~(\ref{T_S}) reflect the sign changes for branch conversion scattering, i.e. 
$\hat{\vk}\rightarrow -\hat{\vk}$ is equivalent to $(\theta\rightarrow\pi-\theta,\phi\rightarrow\pi+\phi)$. 
The factors of $\exp(\pm i\phi')$ in the off-diagonal terms reflect the phase winding of the
order parameter. This parametrization reduces the matrix integral equation to a set of coupled 
one-dimensional integral equations for ${t}^m_a(u',u)$,
\ber
&&
{t}_1^m(u',u) 
= t_N^m(u',u)+\int\limits_{-1}^{1}\frac{du''}{2}t_N^m(u',u'')
 \left[
 \GR(\hat{\vk}'',\varepsilon)
 {t}_1^m(u'',u) 
+(-1)^m e^{-i\phi''}
  \FR(\hat{\vk}'',\varepsilon)
  {t}_3^m(-u'',u)
  \right]
\,,
\label{Eq_1}
\\
&&
{t}_3^m(-u',u) 
= - \int\limits_{-1}^{1}\frac{du''}{2}t_N^{m+1}(u',u'')^{*}
    \left[
    \GR(\hat{\vk}'',\varepsilon)
    {t}_3^m(-u'',u) 
+ (-1)^m e^{i\phi''}
   \FR(\hat{\vk}'',\varepsilon)^{*}
    {t}_1^m(u'',u)
   \right]
\,,
\label{Eq_3}
\\
&&
{t}_2^m(u',-u) 
=\int\limits_{-1}^{1}\frac{du''}{2}t_N^{m-1}(u',u'')
   \left[
   \GR(\hat{\vk}'',\varepsilon)
   {t}_2^m(u'',-u) + (-1)^{m+1}e^{-i\phi''}
   \FR(\hat{\vk}'',\varepsilon)
   {t}_4^m(-u'',-u)
  \right]
\,,
\label{Eq_2}
\\
&&
{t}_4^m(-u',-u) 
= -t_N^{m}(u',u)^{*}-\int\limits_{-1}^{1}\frac{du''}{2}t_N^{m}(u',u'')^{*}
    \left[
    \GR(\hat{\vk}'',\varepsilon)
    {t}_4^m(-u'',-u) 
+ (-1)^{m+1} e^{i\phi''}
    \FR(\hat{\vk}'',\varepsilon)^{*}
    {t}_2^m(u'',-u)
\right]
\,,
\label{Eq_4}
\eer
where
\end{widetext}
\ber
\GR(\hat\vk,\varepsilon) &=&  \frac{\varepsilon}{\sqrt{|\Delta(\hat{\vk})|^2 - \varepsilon^2}}-i
\,,
\\
\FR(\hat\vk,\varepsilon) &=&  \frac{\Delta(\hat{\vk})}{\sqrt{|\Delta(\hat{\vk})|^2 - \varepsilon^2}}
\,.
\eer
Note that ${t}_1^m(u',u)$ is the scattering amplitude for quasiparticle excitations
with angular momentum projection $m$, while $t_4^m(u',u)$ is the corresponding 
quasi-hole amplitude. Branch conversion scattering, generated
by the anomalous propagator, $\FR(\hat\vk,\varepsilon)$, is given by the amplitudes $t_2^m(u',-u)$ (hole 
$\rightarrow$ particle) and $t_3^m(-u',u)$ (particle $\rightarrow$ hole). Multiple scattering 
couples these amplitudes in pairs, $\{{t}_1^m,{t}_3^m\}$ and $\{{t}_2^m,{t}_4^m\}$ as indicated in 
Eqs.~(\ref{Eq_1}-\ref{Eq_3}) and Eqs.~(\ref{Eq_2}-\ref{Eq_4}). The sets of equations are solved numerically.
In Sec.~(\ref{sec-Scattering_Theory}) the solution to the $t$-matrix is used to calculate the differential 
cross-section for the scattering of Bogoliubov quasiparticles off the electron bubble, and from that the forces
on electron bubbles moving in response to an external electric field.

\subsection{Local Density of States}

We first use the $t$-matrix to calculate the spectrum of chiral Fermions bound to the electron bubble. The
asymmetry in the spectrum with respect to the orbital quantum number is responsible for the ground state
current and angular momentum bound to the electron bubble. 
The Nambu Green's function determines the local density of states, 
\be\label{eq-LDOS}
N(\vr,E)=-\frac{1}{2\pi}\mathrm{Im}\left\{\mathrm{Tr}
          \left[\widehat{\cG}^{\text{R}}_{\text{S}}(\vr,\vr;E)\right]\right\}
\,,
\ee
where the trace is over both particle-hole and spin space, and $\widehat\cG^{\text{R}}_{\text{S}}(\vr',\vr;E)$ 
is the retarded Green's function in the presence of an electron bubble. 
It is convenient to express $\widehat\cG^{\text{R}}_{\text{S}}(\vr',\vr;E)$ in momentum space, 
\be\label{eq-G-Fourier}
\hspace*{-3mm}
\cG^{\text{R}}_{\text{S}}(\vr',\vr;E) = 
\ns\int\frac{d^3k}{(2\pi)^3}\ns\int\frac{d^3k'}{(2\pi)^3}e^{i(\vk'\cdot\vr'-\vk\cdot\vr)}
       \cG^{\text{R}}_{\text{S}}(\vk',\vk;E)
\,.
\ee
The low-energy Nambu Green's function for quasiparticles and pairs in the presence of the ion potential can
be expressed in terms of the bulk propagator and the quasiparticle-ion $t$-matrix,
\ber
\widehat{\cG}^{\text{R}}_{\text{S}}(\vk',\vk;E) 
&=& 
(2\pi)^3\delta(\vk'-\vk)\,
\widehat{G}^{\text{R}}_{\text{S}}(\vk,E)
\nonumber\\
&+& 
\widehat{G}^{\text{R}}_{\text{S}}(\vk',E)
\widehat{T}^{\text{R}}_{\text{S}}(\vk',\vk;E)
\widehat{G}^{\text{R}}_{\text{S}}(\vk,E)
\,.
\label{eq-Dyson_T-matrix}
\eer
For energies $|E|\ll E_f$ we can evaluate the $t$-matrix and propagators in the quasiclassical 
approximation and obtain an explicit expression for the local density of states (LDOS).

Figure (\ref{Fig_LDOS}) shows the local density of states calculated at the position, $r=30\,k_f^{-1}$ 
and $\vartheta=\pi/2$, i.e. approximately $19$ Fermi wavelengths from the surface of the
electron bubble. The bulk density of states is shown as the dashed line. A van Hove singularity occurs at the 
maximum gap in the bulk excitation spectrum, while the low-energy spectrum results from the 
nodal quasiparticles near $\theta=0,\pi$. 
Multiple scattering, both potential and branch conversion, by the ion and the chiral order parameter 
generates Andreev bound states indexed by the angular momentum channel, $m$, 
and linear momentum, $p_z=p_f\cos\theta$.
The bound states are broadened into low-energy
bands by integration over $p_z$, and then into
resonances by hybridization with the continuum of nodal quasiparticles. There
are $l_{\text{max}} \approx k_f R\approx 12$ sub-gap resonances shown in Fig.~(\ref{Fig_LDOS}).

More detailed spectral information is obtained by resolving the LDOS in angular momentum channels. 
The reduction of the $t$-matrix as a sum over amplitudes with well defined angular momentum projection 
implies a similar chiral decomposition of the LDOS, 
\ber
N(\mathbf{r},E) = N_0(E) + \sum_{m=-\infty}^{\infty}\delta N_m(\mathbf{r},E)
\,,
\label{LDOS_1}
\eer
where $N_0(E)$ is the bulk density of states in superfluid $\mathrm{^3He-A}$
\ber
N_0(E) = N_F\frac{|E|}{\Delta}\ln\left|\frac{|E| + \Delta}{|E| - \Delta}\right|
\,,
\label{LDOS_homogeneous}
\eer
and $\delta N_m(\vr,E)$, obtained from Eqs.~(\ref{eq-LDOS})-(\ref{eq-Dyson_T-matrix}) with the solutions
of Eqs.~(\ref{Eq_1})-(\ref{Eq_4}), is given by
\begin{widetext}
\ber
\hspace*{-5mm}\delta N_m(\mathbf{r},E) 
\ns&=&\ns
4\pi^2N_f\,\mathrm{Im}\Biggl\{\ns\sum_{l,l'=|m|}^{\infty}\Theta_{l'}^{m}(v)\Theta_{l}^{m}(v)
 \ns\int_{-1}^{1}\ns du'\Theta_{l'}^{m}(u')e^{-\sqrt{\Delta^2(1-u'^2)-\varepsilon^2}\frac{r}{\hbar v_f}}
 \int_{-1}^{1}\ns du\,\Theta_{l}^{m}(u)e^{-\sqrt{\Delta^2(1-u^2)-\varepsilon^2}\frac{r}{\hbar v_f}} 
	K_{l'l}^m(u',u,\varepsilon)\ns\Biggr\}
,
\label{LDOS_m}
\eer
where $u=\cos\theta$ and $u'=\cos\theta'$ in momentum space, $\vr=(r,\vartheta,\varphi)$ is the spatial coordiate
in spherical coordinates with $v=\cos\vartheta$, and the matrix elements, $K_{ll'}^{m}$, are given in terms of 
spherical Bessel functions, the intermediate propagator and the elements of the $t$-matrix (see 
App. \ref{appendix-LDOS_kernel} for details leading to Eq. \ref{LDOS_kernel}),
\begin{align}
K_{l'l}^m(u',u,\varepsilon) 
&=i^{l'-l}\Biggl\{\frac{j_{l'}(k_Fr)j_{l}(k_Fr)}{\sqrt{\Delta^2(1-u'^2)-\varepsilon^2}\sqrt{\Delta^2(1-u^2)-\varepsilon^2}}
   \Bigl[\Delta^2\sqrt{1-u'^2}\sqrt{1-u^2}(t^{m-1}_1 + t^{m+1}_4) + \varepsilon^2(t^m_1 + t^m_4)
\notag\\ 
&+ (-1)^m\Delta\varepsilon\sqrt{1-u'^2}(t^{m}_3 - t^{m}_2) + (-1)^m\Delta\varepsilon\sqrt{1-u^2}(t^{m+1}_2 - t^{m-1}_3)
   \Bigr] 
+ n_{l'}(k_Fr)n_{l}(k_Fr)(t^m_1 + t^m_4)
\notag\\
&-\frac{j_{l'}(k_Fr)n_{l}(k_Fr)}{\sqrt{\Delta^2(1-u'^2)-\varepsilon^2}}
   \Bigl[\varepsilon(t^m_1 - t^m_4) + (-1)^m\Delta\sqrt{1-u'^2}(t^{m}_2 + t^{m}_3)
   \Bigr]
\notag\\
&-\frac{n_{l'}(k_Fr)j_{l}(k_Fr)}{\sqrt{\Delta^2(1-u^2)-\varepsilon^2}}
   \Bigl[\varepsilon(t^m_1 - t^m_4) + (-1)^m\Delta\sqrt{1-u^2}(t^{m+1}_2 + t^{m-1}_3)
   \Bigr]
\Biggr\}
\,\equiv\,i^{l'-l}\,\kappa^{m}_{l'l}(u',u,\varepsilon)
\,.
\label{LDOS_kernel}
\end{align}

\begin{figure}[t]
\centering
\includegraphics[width=0.83\columnwidth,keepaspectratio]{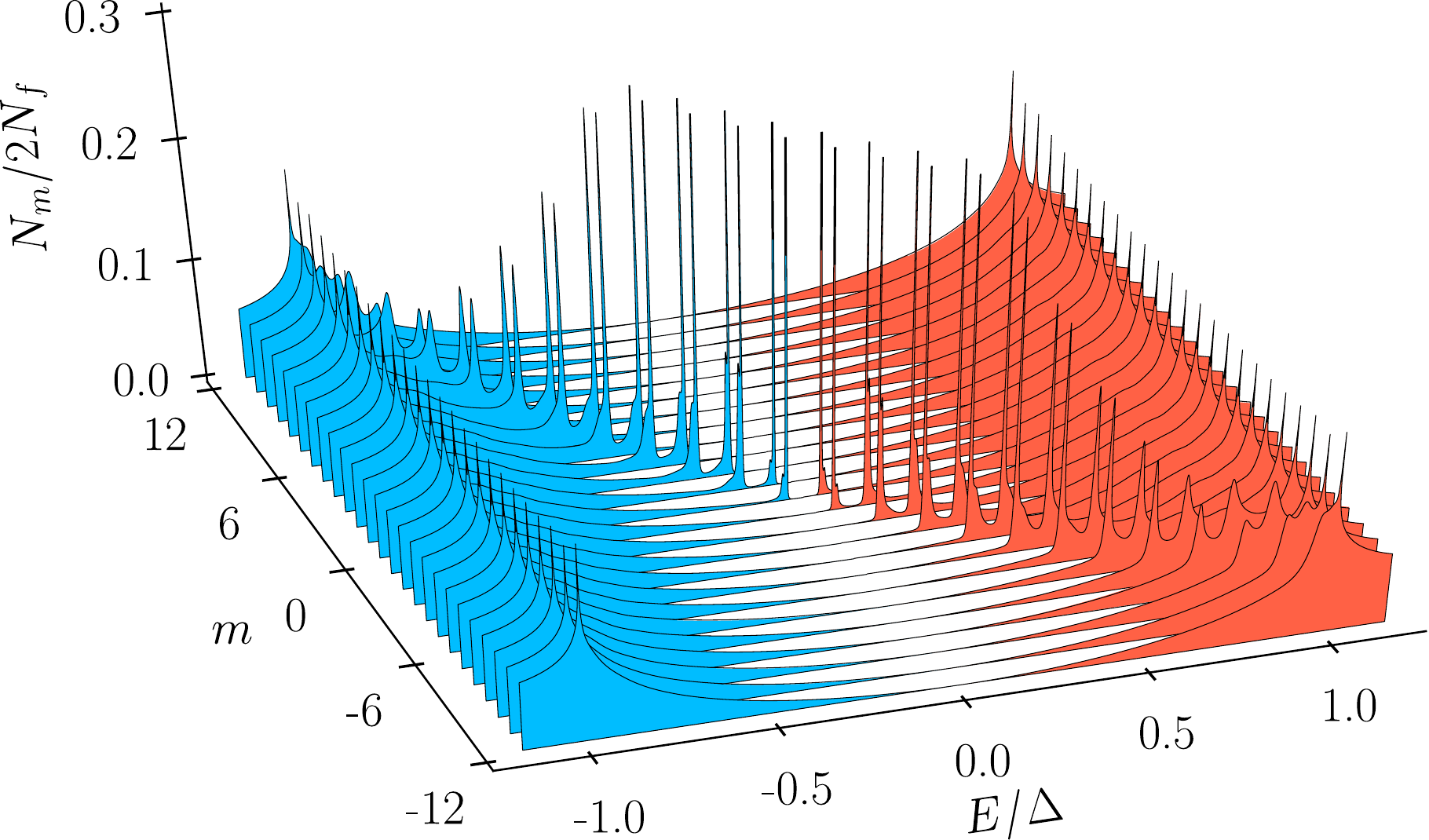}
\caption{
(Color online) The LDOS resolved into orbital angular momentum components labelled by $m$ as 
defined by Eq.~\ref{LDOS_2}. $N_m(\vr,E)$ represents a discrete spectrum of Weyl Fermions 
bound to the electron bubble, plus hybridization with the nodal quasiparticles. 
Note the double degeneracy of each sub-gap energy level.
}
\label{Fig_LDOS_res_m}
\end{figure}
\end{widetext}

The BTRP symmetry of the order parameter $\Delta(\hat{\vk})$ implies that Andreev scattering 
involves transitions between states with $m$ and $m \pm 1$. This mixing of channels is clarified 
by resolving the LDOS in the orbital angular momentum index $m$. 
Here it is worth noting that the sum over $m$ in Eq.~(\ref{LDOS_1}), while formally extending to $\pm\infty$,
is in practice restricted to $|m| \le l_{\text{max}} \approx k_f R\approx 12$ [see 
Fig.~(\ref{figure_hard-sphere_phase-shifts})]. Thus, we resolve the LDOS as
\be
N(\vr,E) = \sum_{m=-l_{\text{max}}}^{l_{\text{max}}}N_m(\vr,E)\,,
\ee
\be
N_m(\vr,E) = \frac{1}{2l_{\text{max}}+1}N_0(E) + \delta N_m(\vr,E)\,.
\label{LDOS_2}
\ee
%
%
Equation (\ref{LDOS_m}) contains exponential factors varying on the coherence length scale, as well as fast 
oscillations, encoded in the spherical Bessel functions, varying on the scale of the Fermi wavelength. 
In Figs.~(\ref{Fig_LDOS}) and (\ref{Fig_LDOS_res_m}) we averaged $\delta N_m(\vr,E)$ over a Fermi wavelength, 
\be\label{QC_average}
\delta N_m^{\mathrm{qc}}(\vr,E) = \frac{1}{\lambda_f}
\int_{r-\lambda_f/2}^{r+\lambda_f/2}\delta N_m(\vr,E)
\,,
\ee
to eliminate the atomic scale oscillations.

In Fig.~(\ref{Fig_LDOS_res_m}) we plot the angular-momentum-resolved LDOS, $N_m(\vr,E)$, as a function 
of energy. Note that the bound states appear in neighboring pairs of $m$-channels, and
that, except for the two states with $m=0$, the bound states for $E\ge 0$ ($E\le 0$) occur only for 
channels with $m > 0$ ($m<0$), the key feature of a Weyl spectrum of chiral Fermions.

\subsection{Bubble Edge Currents}

\begin{figure}[t]
\centering
\includegraphics[width=0.8\columnwidth,keepaspectratio]{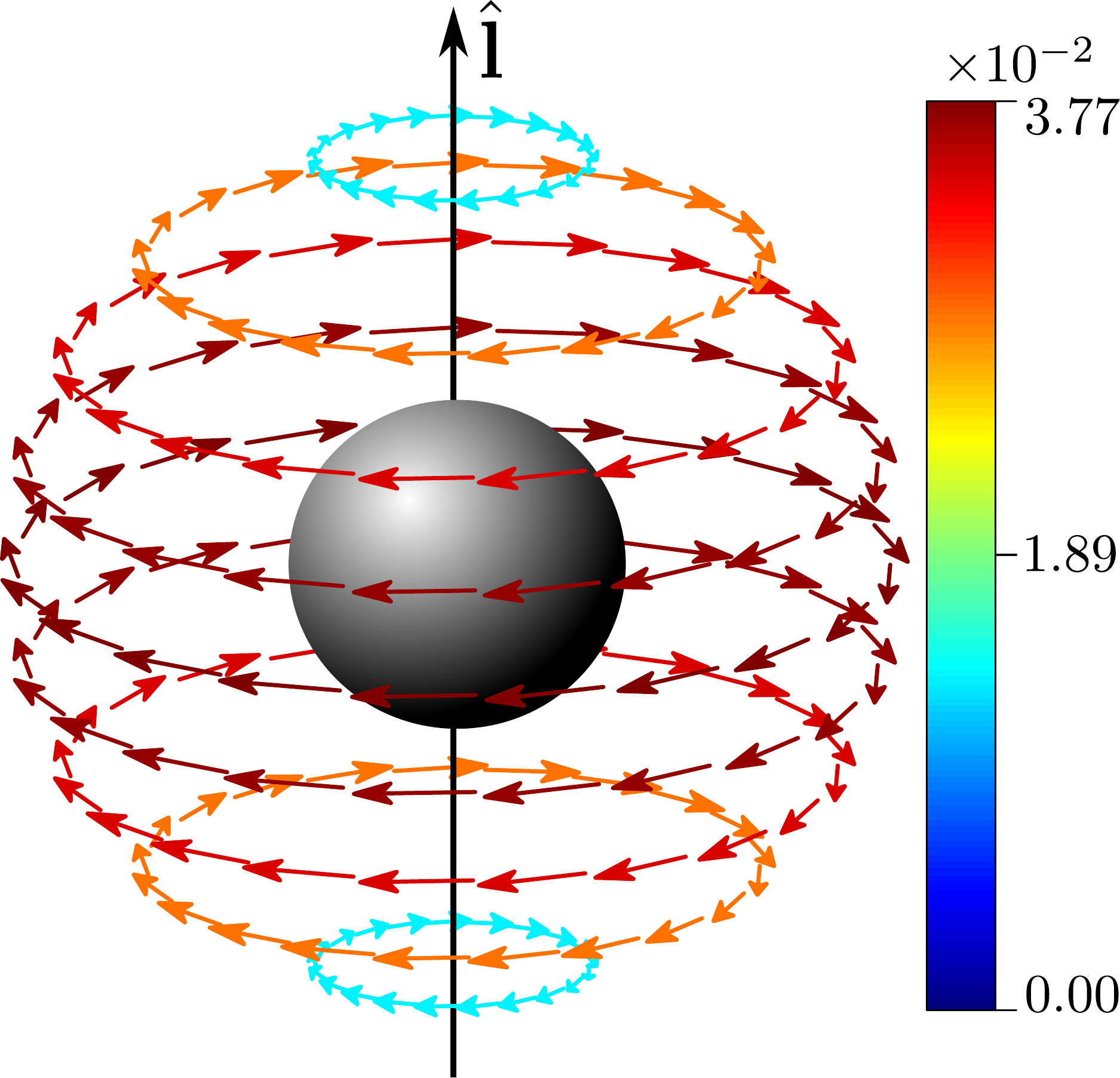}
\caption{
(Color online) Current density $\vj(\vr)$ in units of $N_fv_fk_BT_c$ calculated at 
distance $k_Fr=30.0$ from the bubble center. We used the hard sphere model with $k_FR = 11.17$,
shown in gray. The chiral axis $\hat{\vl}$ determines the direction of Cooper pair angular 
momentum. The temperature is taken as $T = 0.5T_c$.}
\label{Fig_CD_3D}
\end{figure}

The spectrum of chiral Fermions bound to the electron bubble in \Hea\ is responsible for the ground state
current circulating the bubble, the mesoscopic realization of ground-state edge currents on a macroscopic 
boundary of a superfluid \Hea\ film.
The current circulating an electron bubble is calculated from the Fermi distribution and 
the full retarded and advanced Green's functions, 
$\widehat{\cG}^{\text{R,A}}_{\text{S}}(\vr',\vr;E)$, based on 
Eqs.~(\ref{eq-Dyson_T-matrix}), (\ref{eq-GRS_bulk}), (\ref{T_S}) and (\ref{Eq_1})-(\ref{Eq_4}), 
\ber
\hspace*{-3mm}
\vj(\vr) 
&=&
\frac{\hbar}{4}\ns
\int\ns \frac{dE}{2\pi} (f(E)-\nicefrac{1}{2})
\nonumber\\
&\times&
(\boldsymbol{\nabla}_{\vr'}\ns -\ns \boldsymbol{\nabla}_{\vr})
\mathrm{Tr}\,
\left[
\widehat{\cG}^{\text{R}}_{\text{S}}(\vr',\vr;E)
-
\widehat{\cG}^{\text{A}}_{\text{S}}(\vr',\vr;E)
\right]_{\vr=\vr'}
\ns\ns .
\label{eq-GR-Current}
\eer
The current circulating the electron bubble comes from the $t$-matrix term in 
Eq.~(\ref{eq-Dyson_T-matrix}).\footnote{{The propagator for bulk A-phase [Eq.~(\ref{eq-GRS_bulk})] gives 
zero current density, and thus zero angular momentum 
density, when evaluated in the quasiclassical limit with particle-hole symmetry of the normal-state spectrum. 
This is consistent with earlier calculations for the bulk angular momentum density for uniform superfluid \Hea, 
c.f. Ref. \onlinecite{vol81}.}}
%
%
To calculate the current it is more efficient to recast Eq.~(\ref{eq-GR-Current}) in terms of the Matsubara 
Green's function,
\be
\vj(\vr) 
= 
\frac{\hbar}{4i}k_{\text{B}}T\sum_{n=-\infty}^{\infty}
\left[
(\boldsymbol{\nabla}_{\vr'} - \boldsymbol{\nabla}_{\vr})
\mathrm{Tr}\,
\widehat{\cG}^{\text{M}}(\vr',\vr;\epsilon_n)
\right]_{\vr=\vr'}
\,,
\label{eq-GM-Current}
\ee
where $\epsilon_n = (2n+1)\pi k_{\text{B}}T$ are Matsubara frequencies, 
and the Matsubara Green's function is related to the retarded and advanced Green's functions 
by analytic continuation,\cite{AGD2}
\be
\cG^{\text{R(A)}}_{\text{S}}(\vr',\vr,E) = \cG^{\text{M}}_{\text{S}}(\vr',\vr,\epsilon_n)
\Bigr|_{i\epsilon_n\rightarrow E\pm i0},\;\mbox{for $\varepsilon_n \gtrless 0$}
\,.
\ee
After Fourier-transforming Eq.~(\ref{eq-GM-Current}) we obtain
\be
\vj(\vr) 
\ns=\ns 
\frac{\hbar}{4}k_{\text{B}}T\ns\sum_{n=-\infty}^{\infty}\ns
\left[
\sum_{\vk,\vk'}(\vk'+\vk)\,e^{i(\vk'-\vk)\cdot\vr}\,
\mathrm{Tr}\,
\widehat{\cG}^{\text{M}}(\vk',\vk;\epsilon_n)
\right]
\,.
\label{eq-GM-Current2}
\ee
We calculate the current from the propagator, $\widehat{\cG}^{\text{M}}_{\text{S}}(\vk',\vk;\epsilon_n)$, 
by analityic contiuation of the $t$-matrix, 
$\widehat{T}^{\text{M}}_{\text{S}}(\vr',\vr;\epsilon_n)$, which satisfies the system 
of Eqs.~(\ref{Eq_1})-(\ref{Eq_4}), with $\varepsilon\rightarrow i\epsilon_n$. The real-valuedness of 
the current is ensured by the symmetry of the Nambu-Matsubara Green's function,
\be
\widehat{\cG}^{\text{M}}_{\text{S}}(\vk,\vk',-\epsilon_n) 
= 
\left[
\widehat{\cG}^{\text{M}}_{\text{S}}(\vk,\vk',\epsilon_n)
\right]^{\dagger}
\,,
\ee
which also allows us to express the result for the current as a sum over $\varepsilon_n >0$.
The current is purely azimuthal, 
$\vj(r,\vartheta,\phi) = j_{\phi}(r,\vartheta)\ve_{\phi}$ (see App.~(\ref{appendix-Current})), 
with $j_{\phi}(r,\vartheta)$ given by 
\begin{widetext}
\ber
\hspace*{-4mm}
j_{\phi}(r,\vartheta) 
&=& 
-8\pi^3v_fN_fk_BT\,\sum_{n=0}^{\infty}\sum_{m=-\infty}^{\infty}\sum_{l'=|m-1|}^{\infty}\sum_{l=|m|}^{\infty}
J_{l'l}^{m}\,\, 
\Theta_{l'}^{m-1}(\cos\vartheta)\Theta_{l}^{m}(\cos\vartheta)
\,,
\\
J_{l'l}^{m} 
&\equiv& 
\mathrm{Im}\left[i^{l'-l}\right]
\int_{-1}^{1}du'\int_{-1}^{1}du \sqrt{1-u'^2}\,
\Theta_{l'}^{m-1}(u')
e^{-\sqrt{\Delta^2(1-u'^2)+\epsilon_n^2}{\frac{\text{\normalsize $r$}}{\text{\normalsize $\hbar v_f$}}}}
\Theta_{l}^{m}(u)    
e^{-\sqrt{\Delta^2(1-u^2) +\epsilon_n^2}{\frac{\text{\normalsize $r$}}{\text{\normalsize $\hbar v_f$}}}}
\mathrm{Re}\left[\kappa_{l'l}^m(u',u,i\epsilon_n)\right]
\,,
\eer
\end{widetext}
where the functions $\kappa_{l'l}^{m}(u',u,i\varepsilon_n)$ are analytically continued to the 
Matsubara energies, $\varepsilon\rightarrow i\varepsilon_n$.

The current circulating the electron bubble is shown in Fig.~(\ref{Fig_CD_3D}) for 
angular positions on a sphere of radius $r=30\,k_f^{-1}\approx 3.82\,\mbox{nm}$, i.e.
in the near vicinity of the electron bubble with hard sphere 
radius $R=11.17\,k_f^{-1}\approx 1.42\,\mbox{nm}$. 
Note that the axial current flow is opposite to the chiralty of the ground state Cooper pairs for 
all polar angles, and the current vanishes in the direction of the chiral axis, i.e. along the 
nodal points of the order parameter.
The direction of the current flow about the bubble agrees with our expectation based on the direction 
of the edge current for a macroscopic hole, i.e. for a locally translational invariant boundary,
as illustrated in Fig.~(\ref{figure-film_hole}).

\begin{figure}[t]
\centering
\includegraphics[width=0.99\columnwidth,keepaspectratio]{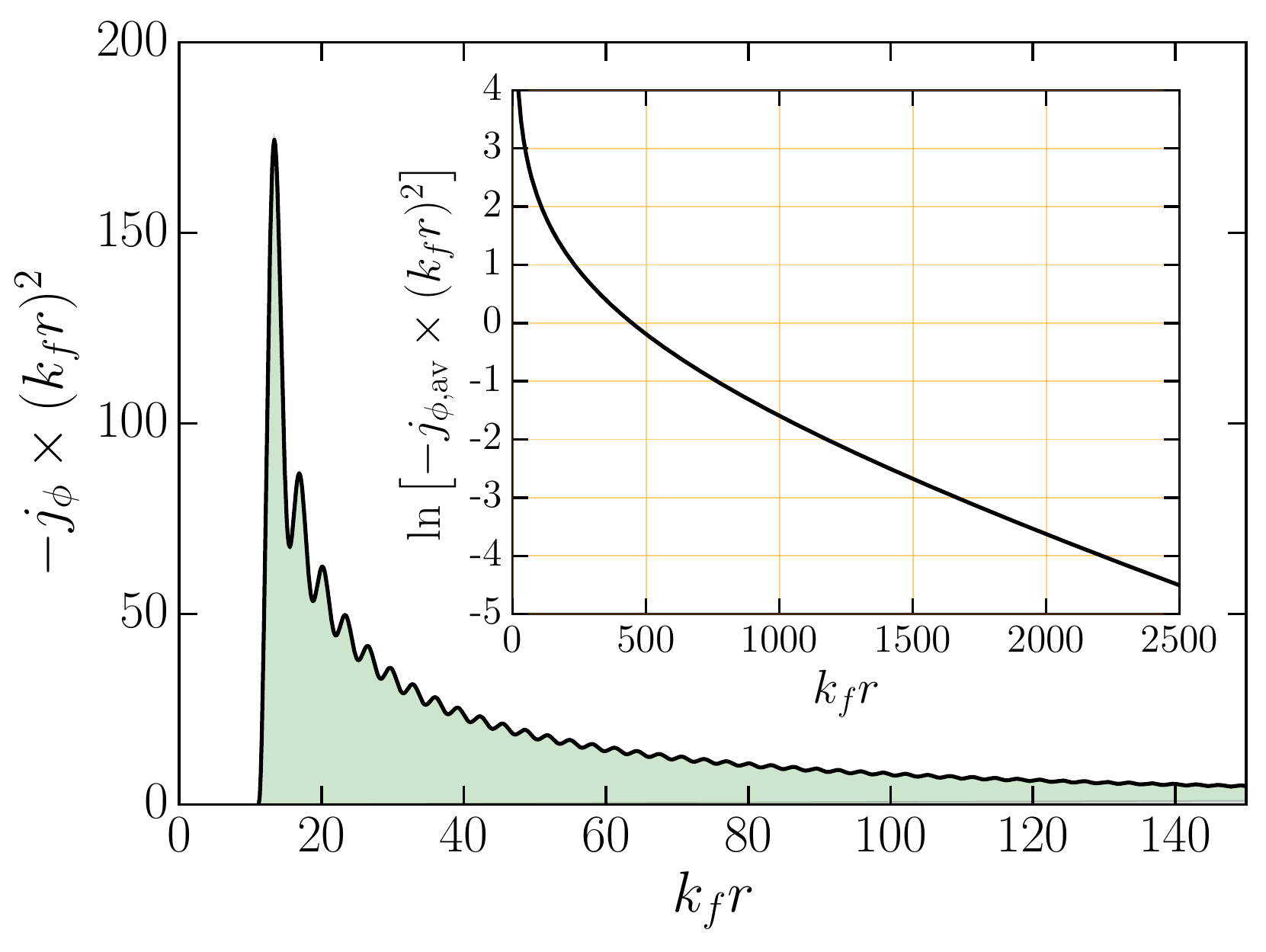}
\caption{
(Color online) 
Radial dependence of the current density $j_{\phi}(r,\vartheta)$ for $\vartheta = \pi/2$,
in units of $v_f N_f k_B T_c$, calculated for the electron bubble with hard sphere radius of 
$k_f R = 11.17$ and temperature $T=0.5T_c$.
The current develops sharply from the bubble radius, then decays rapidly for $r > R$. Quantum 
oscillations on the Fermi wavelength scale are evident at short distances. 
Inset: The current decays exponentially, $j_{\phi,\text{av}}\sim -e^{-r/\xi_0}/(k_f r)^2$, at distances 
greater than the coherence length, $\xi_0=608.7\,k_f^{-1}$; $j_{\phi,\text{av}}$ is 
the quasiclassical envelope obtained by averaging $j_{\phi}(r)$ over a Fermi wavelength as
in Eq. \ref{QC_average}.
}
\label{fig-current_vs_r}
\end{figure}

The current density varies with radial distance from the edge of the electron bubble as shown 
in Fig.~(\ref{fig-current_vs_r}). Note that the current is large on mesoscopic length scales,
$R < r \ll \xi_0$, and decays very rapidly for $k_f r\gtrsim 15$. Quantum oscillations on the scale
of the Fermi wavelength are evident at short distances. For $r \gtrsim \xi_0$ the current density is
small and continues to decay exponentially on the scale of the coherence length as shown in the
inset of Fig.~(\ref{fig-current_vs_r}).

The confinement of the current near the edge of the bubble endows the electron bubble with an angular 
momentum obtained by integrating the angular momentum density from the circulating edge current,
$\vL=\int d^3r\,\vr\times\vj(\vr) = L_z\,\ve_z$,
\be\label{eq-Lz}
L_z= 2\pi\int_{R}^{\infty}r^2 dr\int_{-1}^{+1}d(\cos\vartheta)\,(r\sin\vartheta)\,
     j_{\phi}(r,\vartheta)
\,,
\ee
where the lower limit is set by the vanishing of $j_{\phi}$ for $r\le R$.

\begin{figure}[t]
\centering
\includegraphics[width=0.99\columnwidth,keepaspectratio]{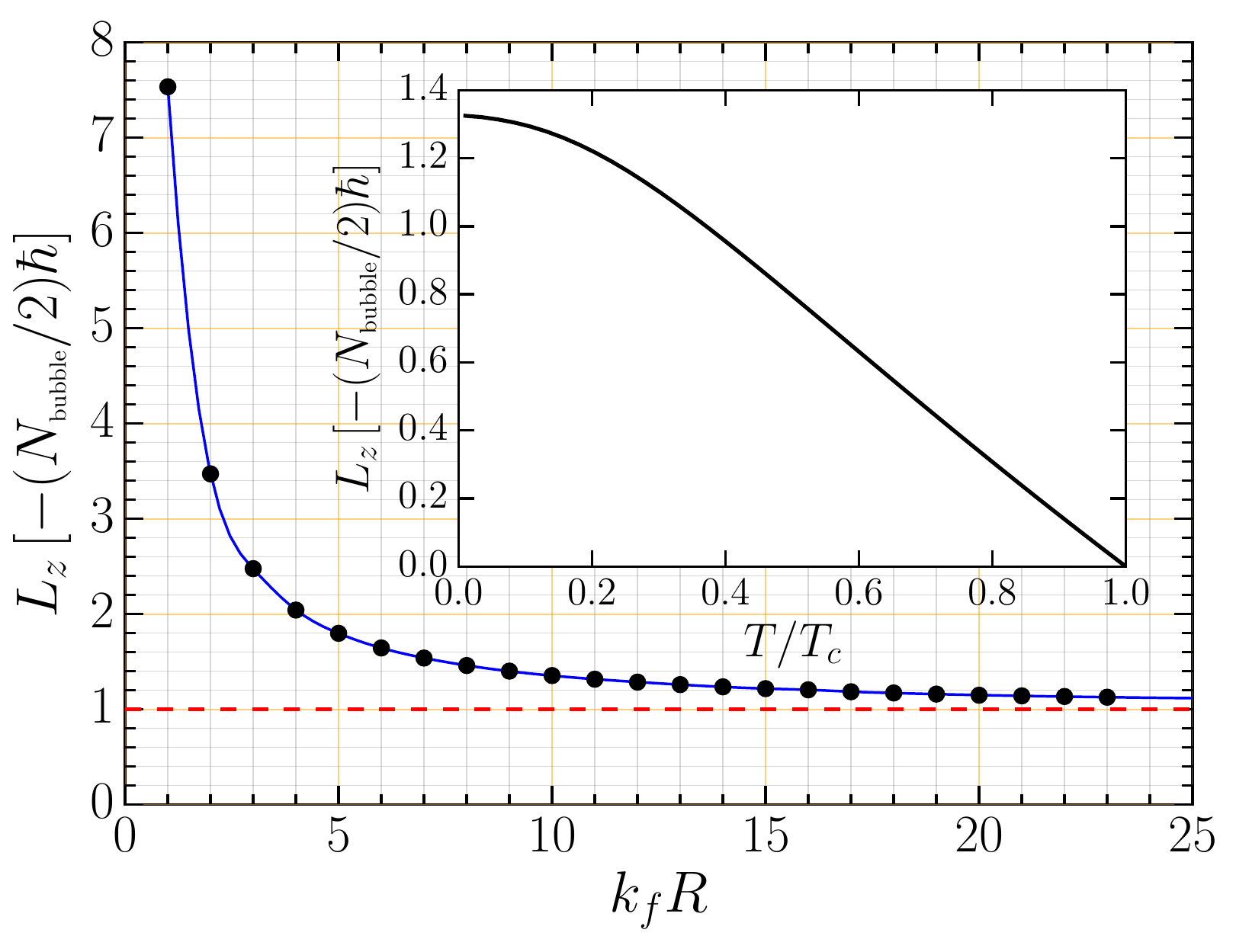}
\caption{
(Color online) 
Angular momentum of hard sphere ``bubbles'' embedded in \Hea\ as a function of  
hard sphere radius, $k_f R$. 
For an electron bubble, $k_f R = 11.17$.
In the macroscopic limit, $R\gg\xi_0$ (red dashed line), 
$L_z$ scales to $L_z^{\infty}=-(N_{\text{bubble}}/2)\,\hbar$
Inset: Temperature dependence of $L_z$ for the electron bubble. 
}
\label{fig-Lz_vs_kfR}
\end{figure}

Recalling our result for the angular momentum generated by edge currents circulating a 
macroscopic hole of radius $R\gg \xi_0$ in a thin \Hea\ film, we express the angular 
momentum of the electron bubble edge currents in similar units, i.e.
\ber
L_z &=& -\mathfrak{f}\,\left(\frac{N_{\text{bubble}}}{2}\right)\,\hbar
\,,
\\
N_{\text{bubble}} &=& \nicefrac{4}{9\pi}(k_f R)^3\approx 197
\,,
\eer
where $N_{\text{bubble}}$ is the number of \He\ atoms excluded from the electron bubble. 
The negative sign reflects 
the fact that the angular momentum of the chiral currents is opposite to the chirality 
of the Cooper pairs.
Numerical integration of Eq.~(\ref{eq-Lz}) gives $\mathfrak{f} = 1.3$, 
remarkably close to the prediction based on the volume of a macroscopic 
hole ($R\gg\xi_0$) in a \Hea\ film, even though the electron bubble is in 
the limit, $R \ll \xi_0$.
Indeed the angular momentum calculated for mesoscopic hard sphere bubbles, scaled in units 
of $-(N_{\text{bubble}}/2)\hbar$, is shown in Fig.~(\ref{fig-Lz_vs_kfR}) 
to rapidly approach the macroscopic scaling result for $k_f R\gg 1$. 
Already at $k_f R = 25$, which corresponds to $R/\xi_0 \approx 0.04$, the
deviation from the macroscopic scaling result is only $\approx 7\%$.
The inset of Fig.~(\ref{fig-Lz_vs_kfR}) shows the temperature dependence of $L_z$ 
for the electron bubble, scaling as $|\Delta(T)|^2\sim|T-T_c|$ in the 
Ginzburg-Landau (GL) limit.\footnote{
This result is at odds with the GL theory result of Rainer and Vuorio,\cite{rai77} who found the 
circulating currents generated by an impurity in \Hea, but with zero net angular momentum.
Their GL calculation for the current and angular momentum is restricted 
to the asymptotic region, $r \gg \xi_0$, where we find the current density is $4-5$ orders of 
magnitude smaller than that in the mesoscopic region $R < r \lesssim \xi_0$.
}

\section{Electron mobility in \Hea}\label{sec-Scattering_Theory}

Application of a d.c. electric field accelerates the electron bubble to a terminal velocity 
$\vv = \dyadic{\mu}\cdot\vE$, where the mobility, $\dyadic{\mu}$,  is determined by 
forces acting on the moving
electron bubble. At finite temperature the mobility is limited by the ``wind'' of
thermal quasiparticles scattering off the moving electron bubble.
In the normal phase of \He\ the scattering rate is sufficiently large that 
recoil of the ion is suppressed, implying elastic scattering and a normal-state 
mobility that is temperature independent.\cite{jos69,and68}
Below $T_c$ the opening of a gap in the excitation spectrum leads to a rapid increase in the         
mobility.\cite{bow77} Experimentally, the mobility increases faster than expected based just 
on the reduction in the number of thermal quasiparticles. Baym et al. \cite{bay77} showed 
that in the superfluid 
B-phase the transport cross-section is also reduced by resonant forward scattering of Bogoliubov 
quasiparticles off the electron bubble. Their theory provides quantitative agreement with 
measurements of the mobility in \Heb\ in the temperature regime near $T_c$.\cite{aho76a}

For the chiral A phase these two basic features also operate. However, superfluid \Hea\
has an anisotropic excitation gap that vanishes for momenta $\vk || \pm \hat\vl$ and
is maximal for momenta $\vk\perp\hat\vl$. Thus, an electron bubble will experience a stronger
drag force for $\vv || \pm\hat\vl$ compared to $\vv\perp\hat\vl$, i.e. $\mu_{||} < \mu_{\perp}$.
Indeed the anisotropy of the negative ion mobility was calculated by extending the 
scattering theory for the B-phase by Baym et al.\cite{bay77} to scattering by an ion 
in \Hea,\cite{sal80,sal87b} and measurements of the mobility anisotropy, 
$\mu_{\perp}-\mu_{||}$ were made via pulse-shape, time-of-flight experiments on vortex 
textures of superfluid \Hea.\cite{sim86} 
Note that the drag force on the electron bubble is insensitive to the direction of the 
chiral axis, i.e. the drag force for $+\vE || \hat\vl$ and $-\vE || \hat\vl$ are the same. 

The chiral axis is a reflection of broken time-reversal symmetry ($\mathrm{T}$)
and broken mirror symmetry ($\Pi$) in a plane containing the chiral axis 
$\hat\vl$. The generalization of the mobility for the isotropic B-phase to \Hea\ with 
chiral axis $\hat\vl || \hat\vz$ is a mobility tensor, $\mu_{ij}$ with $i,j\in\{x,y.z\}$; 
thus, $v_i=\mu_{ij}E_j$, where the components are all real.
Uniaxial rotation symmetry restricts the elements of the mobility tensor to 
$\mu_{xx}=\mu_{yy} \equiv \mu_{\perp}$, $\mu_{zz}\equiv\mu_{||}$, and $\mu_{xy}=-\mu_{yx}$; 
all other components vanish. Thus, the electron mobility tensor for \Hea\ has the form 
\be
\mu_{ij} = \mu_{\perp}(\delta_{ij}-\hat\vl_i\hat\vl_j) 
         + \mu_{||}\hat\vl_i\hat\vl_j
         + \mu_{xy}\epsilon_{ijk}\hat\vl_k
\,.
\ee

The off-diagonal component, $\mu_{xy}$, is allowed by axial rotation symmetry and chiral 
symmetry, $\mathrm{C}=\mathrm{T}\times\Pi$, but vanishes if the ground state is separately 
invariant under mirror symmetry, $\Pi$, in a plane containing the chiral axis $\hat\vl$. 
This would be the case for a Planar phase of \He, which is 
degenerate in weak-coupling theory with the A-phase, has the same anisotropic excitation 
gap, and thus, is indistinguishable from \Hea\ in terms of $\mu_{||}$ and $\mu_{\perp}$.
What distinguishes the A-phase is that neither $\mathrm{T}$ nor $\Pi$ are symmetries.
The breaking of both $\mathrm{T}$ and $\Pi$ allows for $\mu_{xy}\ne 0$, and thus
transverse motion of the electron bubble for $\vE\perp\hat\vl$, i.e. an anomalous Hall current
of electron bubbles given by
\be
\vv_{\text{AH}} = \mu_{xy}\,\vE\times\hat{\vl}
\,.
\ee
More generally, for any field orientation, the steady state ion velocity is given by 
\be\label{eq-velocity_vs_field}
\vv = \mu_{\perp} \hat\vl\times(\vE\times\hat\vl) 
    + \mu_{||}(\hat\vl\cdot\vE)\,\hat\vl 
    + \mu_{xy}\vE\times\hat\vl
\,.
\ee

This steady state result for the velocity arises from the balance between the Coulomb
force, $\vF_{\tiny \vE}= e\vE$, and the quasiparticle force, 
$\vF_{\text{QP}} = -\dyadic{\eta}\,\vv$, where $\dyadic{\eta}$ is the generalized Stokes
tensor for an anisotropic fluid.
The latter determines the inverse of the mobility tensor, 
$\dyadic{\eta} = e\dyadic{\mu}^{-1}$, and has the same structure as the mobility tensor, 
$\eta_{ij} = \eta_{\perp}(\delta_{ij}-\hat\vl_i\hat\vl_j) 
          + \eta_{||}\hat\vl_i\hat\vl_j
          + \eta_{xy}\epsilon_{ijk}\hat\vl_k
$.
Theoretically, we determine the force on a moving ion, i.e. the Stokes tensor. The
components of the mobility are then given by the inversion formulas,
\be
\mu_{\perp} = e\frac{\eta_{\perp}}{\eta_{\perp}^2 + \eta_{xy}^2}
,\quad
\mu_{xy}    = e\frac{-\eta_{xy}}{\eta_{\perp}^2 + \eta_{xy}^2}
,\quad
\mu_{||}    = e\frac{1}{\eta_{||}}
\,.
\label{mobility-Stokes}
\ee

\subsection{Quasiparticles Forces on an Electron Bubble}

We formulate the microscopic theory for the forces acting on a moving electron bubble due to 
scattering by thermal quasiparticles in the chiral A phase of \He. The key assumptions are 
(i) that the velocity of the electron bubble is sufficiently low that the resulting Stokes 
tensor, $\eta_{ij}$, is independent of the electron velocity, (ii) 
the recoil energy of the ion is sufficiently low, $\Delta_{\text{rec}}\ll k_BT$,
that it is a good approximation to consider quasiparticle-ion scattering in the elastic
limit, (iii) the ground state is described by the ESP chiral A phase order 
parameter in Eq.~(\ref{BdG}) and (iv) the only input parameters to the theory are the normal state
scattering phase shifts constrained by the normal state mobility 
[Eq.~(\ref{delta_l_hs}) and Fig.~(\ref{figure_hard-sphere_phase-shifts})].

Our analysis is close to that of Baym et al. for the ion mobility in \Heb,\cite{bay77,bay79} except 
that we incorporate broken time-reversal and mirror symmetries of the chiral ground state 
into the theory of the transport cross-section for scattering of Bogoliubov quasiparticles
off the electron bubble embedded in \Hea. 
Earlier theoretical analyses of the electron mobility in \Hea\ included the ansisotropy of the 
excitation spectrum, but imposed mirror symmetry in the formulation of
the scattering of Bogoliubov quasiparticles off the ion embedded in \Hea.\cite{sal80,sal89,sal90} 
See App.~(\ref{appendix_Salmelin-errors}) for our critique of earlier work.

In what follows we derive results for the scattering cross section and forces on a 
negative ion moving in superfluid \Hea\ driven by a static electric field.
We start from the equation of motion for the momentum of the ion,
\be\label{dpdt_1}
\frac{d\mathbf{P}}{dt} 
=
- \sum_{\vk,\vk'}\hbar(\vk'-\vk)(1-f_{\vk'}) f_{\vk}\,\,\Gamma_{\vv}(\vk',\vk)
\,,
\ee
where $\hbar(\vk-\vk')$ is the momentum transferred to the ion by scattering of a 
quasiparticle from $\vk\rightarrow\vk'$, $f_{\vk} $ is the  probability that the 
incident state $\vk$ is occupied, $(1-f_{\vk'})$ is the probability that the final 
state $\vk'$ is unoccupied, and $\Gamma_{\mathbf{v}}(\vk',\vk)$ is the transition 
rate of scattering of quasiparticles by the ion moving with velocity $\vv$. 
In the low velocity limit the forces are linear in $\vv$. Generalization of the 
theory presented here to higher velocities when inelastic scattering and non-linear 
velocity dependence becomes important is outside the scope of this report, but 
can be formulated as a generalization of the theory of Josephson and Lekner for 
the dynamics of electrons in normal \He.\cite{jos69}

In the low velocity limit the motion of an ion does not substantially perturb the 
initial and final quasiparticle distribution functions, i.e. the ion moves through 
a Fermi-Dirac distribution of quasiparticles described by temperature $T$, 
$f_{\vk} = f(E_{\vk}) \equiv [1+\exp(E_{\vk}/k_BT)]^{-1}$, where 
$E_{\vk} = \sqrt{\xi_{\vk}^2 + |\Delta(\vk)|^2}$ is the bulk \Hea\ excitation energy.

To linearize Eq.~(\ref{dpdt_1}) in the ion velocity we follow Baym et al.\cite{bay69} 
and observe that if the distribution of quasiparticles were in thermal equilibrium 
and co-moving with the electron bubble, then the initial and final state distribution 
functions would be Doppler-shifted Fermi-Dirac distributions,
\be
\bar{f}_{\vk} = f(E_{\vk}-\hbar\vk\cdot\vv)
\,.
\ee

In this case the net momentum transfer is zero. We then subtract zero from 
Eq.~(\ref{dpdt_1}) to obtain,

\be\label{dpdt_2}
\hspace*{-3mm}
\frac{d\mathbf{P}}{dt} = - \sum_{\vk,\vk'}\hbar(\vk'-\vk)\!\left[f_{\vk}(1-f_{\vk'}) 
                         - \bar{f}_{\vk}(1-\bar{f}_{\vk'})\right]\!\Gamma_{\vv}(\vk',\vk)
\,.
\ee

The momentum transfer to the ion is a sum over all incident and final state momenta.
For every transition, $\vk\rightarrow\vk'$, there is a mirror scattering event, 
$\vk'\rightarrow\vk$, that contributes to the net transfer of momentum to the ion. 
In order to isolate the scattering events responsibile for the anomalous Hall mobility it is convenient 
to symmetrize the right-hand side of Eq.~(\ref{dpdt_2}) and express the momentum transfer rate in 
terms of pairs of transition rates related by mirror symmetry,

\begin{widetext}
\ber
\frac{d\mathbf{P}}{dt} 
= -\nicefrac{1}{2}\sum_{\vk,\vk'}\hbar(\vk'-\vk)
\Bigl\lbrace\left[f_{\vk}(1-f_{\vk'})-\bar{f}_{\vk}(1-\bar{f}_{\vk'})\right]\Gamma_{\vv}(\vk',\vk)
-
\left[f_{\vk'}(1-f_{\vk})-\bar{f}_{\vk'}(1-\bar{f}_{\vk})\right]\Gamma_{\vv}(\vk,\vk')\Bigr\rbrace
\,.
\label{dpdt_3}
\eer
\end{widetext}

A key point is that the phase space factors for allowed transitions - the terms in square brackets - 
are already linear in the ion velocity $\vv$. Thus, we evaluate the transition rate, $\Gamma_{\vv}$, in 
the static limit, $\Gamma_{\vv}(\vk',\vk)\rightarrow \Gamma(\vk',\vk)$, with the latter given by 
Fermi's golden rule,

\vspace*{1mm}
\ber
\Gamma(\vk',\vk) &=& \frac{2\pi}{\hbar} W(\vk',\vk) \delta(E_{\vk'}-E_{\vk})
\label{Fermi_GR}
\,,
\eer
where $W(\vk',\vk)$ is the transition rate for Bogoliubov quasiparticles, defined by the 
Bogoliubov-Nambu spinors in Eqs.~(\ref{BN_spinor1})-\ref{BN_spinor2}, scattering off
the electron bubble,

\ber
\label{Scattering_Rate}
W(\vk',\vk) 
= 
\frac{1}{2}
\sum_{\sigma,\sigma' = \uparrow,\downarrow}
&\Big\{&
|\bra{\Psi_{1,\vk'\sigma'}}\widehat{T}_S\ket{\Psi_{1,\vk\sigma}}|^2
\\
&+& 
|\bra{\Psi_{1,\vk'\sigma'}}\widehat{T}_S\ket{\Psi_{2,\vk\sigma}}|^2 
\nonumber\\
&+& 
|\bra{\Psi_{2,\vk'\sigma'}}\widehat{T}_S\ket{\Psi_{1,\vk\sigma}}|^2 
\nonumber\\
&+&
|\bra{\Psi_{2,\vk'\sigma'}}\widehat{T}_S\ket{\Psi_{2,\vk\sigma}}|^2
\Big\}_{E_{\vk'}=E_{\vk}}
\,.
\nonumber
\eer

The result for the scattering rate for Bogoliubov quasiparticles is a sum over the possible elastic 
scattering events between Bogoliubov particle-like (1) and hole-like (2) branches of the excitation 
spectrum: 
$1\rightarrow 1$,
$2\rightarrow 1$,
$1\rightarrow 2$,
and 
$2\rightarrow 2$.
Expanding the Doppler-shifted Fermi functions in Eq.~(\ref{dpdt_3}) to linear order in $\vv$ yields

\begin{widetext}
\ber
\frac{d\mathbf{P}}{dt} 
&=& -\nicefrac{3}{2}
    \hbar k_f n_3\,k_f^{-2}
    \left(\frac{m^{\ast}}{2\pi\hbar^2}\right)^2
    \int d\Omega_{\vk}
    \int\frac{d\Omega_{\vk'}}{4\pi}
    \int_{|\Delta({\vk})|}^{\infty}dE
    \int_{|\Delta({\vk}')|}^{\infty}dE'
    \,\delta(E-E')
    \frac{EE'}{\sqrt{E^2-|\Delta({\vk})|^2}\sqrt{E'^2-|\Delta({\vk}')|^2}}
    \left(-\frac{\partial f}{\partial E}\right)
\nonumber\\
&\times&
    \,({\vk}'-{\vk})
    \Big\{
    W(\vk',\vk)\,
    \left[\vk' f - \vk (1-f)\right]
     -
    W(\vk,\vk')\,
    \left[{\vk}f - {\vk}'(1-f)\right]
    \Big\}
    \cdot\vv
\,,
\label{dpdt_4}
\eer
\end{widetext}
where $f=[\exp(E/k_BT)+1]^{-1}$ and we used the fact that the momenta are restricted to, $|k-k_f|\ll k_f$, 
and energies are confined to a shell near the Fermi surface, $|\xi_{\vk}| =v_f|(k-k_f|\ll E_f$.
We changed energy integration variables from $d\xi_{\vk}\rightarrow dE_{\vk}$ with 
$\xi_{\vk}=\pm\sqrt{E_{\vk}^2 -|\Delta(\hat\vk)|^2}$, where $\xi_k > 0$ and $\xi_k < 0$ correspond to 
particle-like and hole-like excitations, respectively. In Eq.~(\ref{dpdt_4}) and hereafter, the momenta 
are evaluated on the Fermi surface: $\vk=k_f\hat\vk$ and $\vk'=k_f\hat\vk'$, and 
$W(\vk',\vk)=W(\hat\vk',\hat\vk;E)$.

\subsection{Microscopic Reversibility \& Mirror Symmetry}\label{sec-Microscopic_Reversibility}

If the ground state in which the ion is embedded were time-reversal and mirror symmetric we could use 
the ``microscopic reversibility'' condition, $W(\vk',\vk) = W(\vk,\vk')$.
This is the case for the B-phase of \He, which also has a rotational invariant excitation spectrum
and bulk gap, $|\Delta(\hat\vk)|=\Delta$. Equation (\ref{dpdt_4}) then reduces to $d\vP/dt = - \eta\,\vv$, 
with the Stokes drag coefficient, and thus the inverse mobility, given by 
\ber
\eta  = \frac{e}{\mu} = n_3\,p_f\,\int_{\Delta}^{\infty}dE\,\sigma^{\text{tr}}(E)
          \left(-2\frac{\partial f}{\partial E}\right)
\eer
where the energy resolved transport cross-section is 
\ber
\sigma_{\text{tr}}(E) 
=
\left|\frac{m^*}{2\pi\hbar^2}\der{\xi}{E}\right|^2
\int d\Omega_{\vk'}
W_{\text{B}}(\hat\vk'\cdot\hat\vk;E)\, (1-\hat\vk'\cdot\hat\vk)
\,,\qquad
\eer
in agreement with the result for the mobility obtained for \Heb\ by Baym et al.\cite{bay77}
In the limit $\Delta\rightarrow0$ this result reduces to the mobility of normal \He\ given by 
Eq.~(\ref{eq-mobility_normal-state}).

The theory for the mobility of \Heb\ was extended by Salomaa et al. to calculate the mobility 
tensor \Hea.\cite{sal80} These authors included the anisotropy of the excitation gap, $|\Delta(\hat\vk)|$. 
However, they implicitly assumed mirror symmetry by imposing the microscopic 
reversibility condition for mirror symmetric scattering events.
Microscopic reversibility implies that the second line of Eq.~(\ref{dpdt_4}) reduces to 
$\times\,({\vk}'-{\vk})\, W(\vk',\vk)\,({\vk}'-{\vk})\cdot\vv$. The resulting momentum transfer to the 
ion by quasiparticle scattering is then given by a \emph{symmetric} Stokes tensor, and thus there is 
no transverse force on the moving ion. Indeed in Ref.~[\onlinecite{sal80}] the uniaxial anisotropy of
the mobility tensor was calculated, but no anomalous Hall term was reported.

Existence of a transverse force acting on an electron bubble moving in \Hea\ was argued on 
physical grounds by Salmelin et al.\cite{sal89} based on the prediction of currents circulating an 
impurity in superfluid \Hea,\cite{rai77} and the analogy with the Magnus effect arising from
the hydrodynamic lift force on a rotating sphere moving through a fluid.\cite{wat87} The authors 
dubbed the transverse force on a moving ion in \Hea\ an ``intrinsic Magnus effect'', and they focused
their discussion on the limit of a small object such as an electron bubble with radius $R\ll\xi_0$, 
small in comparison to the size of the Cooper pairs in \Hea. 

Although the basic picture motivating the existence of a transverse force on electron bubbles moving 
through a chiral superfluid is sound, the microscopic theory outlined in Ref.~[\onlinecite{sal89}], 
and published in detail by Salmelin and Salomaa in Ref.~[\onlinecite{sal90}], is fundamentally flawed.
These authors impose mirror symmetry in their calculation of the scattering amplitude for 
momentum transfer from the distribution of quasiparticles to the moving ion by adopting the 
microscopic reversibility condition $W(\vk',\vk)=W(\vk,\vk')$.
This equality gaurantees, within scattering theory, that there is no transverse force 
on the electron bubble. As a consequence the theoretical results and prediction for the 
transverse Hall mobility in Refs.~[\onlinecite{sal89,sal90}] are spurious.
We include a more detailed critique of this work in App.~(\ref{appendix_Salmelin-errors}).

In the following section we show that it is precisely the asymmetry in scattering rates
for $\vk\rightarrow\vk'$ and its mirror symmetric partner, $\vk'\rightarrow\vk$, that is the origin of
the transverse force acting on a moving electron bubble.

\subsection{Scattering Cross Sections and the Mobility Tensor}\label{sec-cross-sections}

A central feature of Eq.~(\ref{dpdt_4}) is that the rates $W(\vk',\vk)$ and $W(\vk,\vk')$ for 
mirror symmetric scattering events are \emph{not equal} for chiral ground states like that of 
superfluid \Hea. To highlight the importance of this fact we separate $W(\vk',\vk)$ into its 
mirror symmetric ($W^{+}$) and anti-symmetric ($W^{-}$) parts,

\be\label{W_+-}
W(\vk',\vk) = W^{(+)}(\vk',\vk) + W^{(-)}(\vk',\vk)
\,,
\ee
with $W^{(\pm)}(\vk',\vk) = \pm W^{(\pm)}(\vk,\vk')$.
Equation~(\ref{dpdt_4}) for the force on the moving ion is linear in the ion velocity,
$d\vP/dt=-\dyadic{\eta}\cdot\vv$, and can be expressed in terms of the components of the Stokes tensor, 

\ber
\hspace*{-2mm}
\eta_{ij} = n_3p_f\int_{0}^{\infty} dE\left(-2\frac{\partial f}{\partial E}\right)\sigma_{ij}(E)
\,,\quad i,j\in\{x,y,z\}\,,
\label{eq-Stokes_Tensor}
\eer
where $\sigma_{ij}(E) = \sigma^{(+)}_{ij}(E) + \sigma^{(-)}_{ij}(E)$ 
is the energy-resolved transport cross-section separated into
symmetric ($+$) and anti-symmetric ($-$) tensor components,
$\sigma^{(\pm)}_{ij}(E)$, which are given by Fermi surface averages over the 
differential cross-section,

\begin{widetext}
\ber
&&
\frac{d\sigma}{d\Omega_{\vk'}}(\hat{\vk}',\hat{\vk};E) 
= 
\left(\frac{m^{\ast}}{2\pi\hbar^2}\right)^2
\frac{E}{\sqrt{E^2-|\Delta(\hat{\vk}')|^2}}\,
W(\vk',\vk)\,
\frac{E}{\sqrt{E^2-|\Delta(\hat{\vk})|^2}}
\,,
\label{dsigma-dOmega}
\\
\,&&\,
\nonumber\\
\sigma^{(+)}_{ij}(E) 
&=& 
\frac{3}{4}
\int_{E\geq|\Delta(\hat{\vk}')|} d\Omega_{\vk'}\,
\int_{E\geq|\Delta(\hat{\vk})|}\frac{d\Omega_{\vk}}{4\pi}\,
\,\,\big[
(\hat{\vk}'_i - \hat{\vk}_i)(\hat{\vk}'_j - \hat{\vk}_j)\,
\big]\,\,
\frac{d\sigma}{d\Omega_{\vk'}}(\hat{\vk}',\hat{\vk};E)\,
\,,
\label{sigma_+}
\\
\sigma^{(-)}_{ij}(E) 
&=& 
\frac{3}{4}
\int_{E\geq|\Delta(\hat{\vk}')|} d\Omega_{\vk'}\,
\int_{E\geq|\Delta(\hat{\vk})|}\frac{d\Omega_{\vk}}{4\pi}\,
\,\,\big[
\varepsilon_{ijk}(\hat{\vk}'\times\hat{\vk})_{k}
\big]\,\,
\frac{d\sigma}{d\Omega_{\vk'}}(\hat{\vk}',\hat{\vk};E)\,
\Big[f(E)-\nicefrac{1}{2}\Big]
\,.
\label{sigma_-}
\eer
\end{widetext}

Equations (\ref{eq-Stokes_Tensor})-(\ref{sigma_-}) combined with Eqs.~(\ref{Scattering_Rate}) 
and (\ref{Eq_1}-\ref{Eq_4}) to compute the scatteing rate $W(\vk',\vk)$, are the central 
results for the forces on a moving electron bubble. The Stokes tensor determines both the 
drag forces, $\propto\eta_{\perp,||}$, and the transverse force, $\propto\eta_{xy}$, responsible 
for the anomalous Hall effect on moving electron bubbles in chiral superfluid phase of \He.

Note that Eq.~(\ref{sigma_+}) for the symmetric part of the transport cross section is equivalent
to Eqs. [7] and [8] of Ref.~[\onlinecite{sal90}]. This is a symmetric tensor, and as is clear from
the integrand of Eq.~(\ref{sigma_+}) \emph{only} the symmetric part of the scattering 
rate, $W^{(+)}(\vk',\vk)$, contributes to $\sigma^{(+)}_{ij}(E)$. Thus, $\sigma^{(+)}_{ij}(E)$ 
contributes only to the diagonal components of the Stokes tensor; there is no anomalous Hall 
term contained in Eq.~(\ref{sigma_+}). The errors leading the authors of 
Refs.~[\onlinecite{sal89,sal90}] to obtain $\sigma^{(+)}_{xy}\ne 0$ 
are identified and discussed in App.~(\ref{appendix_Salmelin-errors}).

The anti-symmetric part of the transport cross section given by Eq.~(\ref{sigma_-}) is the origin
of the transverse force on a moving electron bubble. This is a new result that is present
because quasiparticle scattering off an electron bubble embedded in a chiral superfluid 
acquires a spectrum of chiral Fermions bound to the electron bubble. As a result the scattering
rates for $\vk\rightarrow \vk'$ and the mirror-symmetric scattering event, $\vk'\rightarrow \vk$,
are not equal. From the integrand of Eq.~(\ref{sigma_-}) it is clear that \emph{only} the 
anti-symmetric part of the scattering rate, $W^{(-)}(\vk',\vk)$, contributes
to $\sigma^{(-)}_{ij}(E)$. The anti-symmetric cross section, $\sigma^{(-)}_{ij}(E)$,
determines the off-diagonal components of the Stokes tensor, and thus the transverse force
acting on the moving electron bubble.\footnote{The derivation of Eq.~(\ref{sigma_-}) includes a 
term, $\propto\int d\Omega_{\vk'}\int d\Omega_{\vk}(\hat{\vk}'_i\hat{\vk}'_j-\hat{\vk}_i\hat{\vk}_j)
d\sigma^{(-)}/d\Omega_{\vk'}(\hat{\vk}',\hat{\vk};E)$, which, based on the analysis in 
App.~(\ref{App_CSs}) vanishes identically.}
Note that $\sigma^{(-)}_{ij}(E)$ is identically zero 
if the condition of microscopic reversibility is assumed to hold, i.e. $W^{(-)}\equiv 0$.
Also note the distribution function, $f(E)-\nicefrac{1}{2}=-\nicefrac{1}{2}\tanh(E/2k_B T)$,
appearing in Eq.~(\ref{sigma_-})
is odd under $E\rightarrow -E$. This implies that the transverse force originates from the 
chiral part of the spectrum, 
which is a 
reflection of branch conversion scattering between particle-like and hole-like excitations 
by the chiral order parameter.

Lastly, for $|\Delta(\hat{\vk})| = 0$ Eqs.~(\ref{sigma_+}) and (\ref{sigma_-}) reduce to 
the normal-state transport cross-section given in Eq.~(\ref{sigma_N}),
\be
\sigma^{(+)}_{ij}(E) \rightarrow \delta_{ij}\,\sigma^{\text{tr}}_{\text{N}}
\,,\quad
\sigma^{(-)}_{ij}(E) \rightarrow 0
\,,
\ee
where $\sigma^{(-)}_{ij}(E)$ vanishes because the gauge and mirror symmetries are unbroken in the 
normal Fermi liquid. Integration over energy in Eq.~(\ref{eq-Stokes_Tensor}) gives unity and 
we obtain the Stokes drag, and thus the temperature independent normal-state mobility given by 
Eq.~(\ref{eq-mobility_normal-state}).

\section{Results for the e$^{-}$ mobility in \Hea}\label{sec-Results_Electron-Mobility}

Our formulation of the scattering theory was motivated in part by the reports of 
the RIKEN group of an anomalous Hall effect in their measurements of electron transport in superfluid 
\Hea\ for temperatures down to $T\approx 250\,\mu\mbox{K}$.\cite{ike13,ike15} 
In these experiments electrons are forced to a depth of $30\,\mbox{nm}$ below the free surface of 
liquid \He\ by a perpendicular electric field. The electrons form negatively charged bubbles with an 
effective mass $M\approx\nicefrac{2}{9\pi}(k_f R)^3\,m_3\approx 100\,m_3$, where $m_3$ is the mass of 
the \He\ atom.\cite{and68}
For \Hea\ the chiral axis is locked normal to the free surface, $\hat\vl || \hat\vz$.  
The electron bubbles are then driven into motion by an additional electric field 
$\vE=\cE\,\ve_x$ applied in the $xy$-plane. A pair of split electrodes are used to measure 
both the longitudinal current, $v_x=\mu_{xx}\cE$, and the Hall current, 
$v_y=\mu_{xy}\cE$.\footnote{{The experiments are carried out 
at a.c. frequencies from $1-10\,\mbox{Hz}$. The current response contains both an in-phase and 
out-of-phase a.c. components which can be calculated from the hydrodynamical equations with 
the $M$ and $\eta_{ij}$ calculated in the low velocity, d.c. limit.}}

The RIKEN group also compared their measurements of the anomalous Hall angle for electron bubbles 
in superfluid \Hea, with calculations based on the theoretical formulas for the longitudinal
and transverse mobilities published in Ref.~[\onlinecite{sal90}]. 
However, the comparison is based on a fundamentally flawed theory of the mobility tensor,
particularly the anomalous Hall effect [see discussion in App.~(\ref{appendix_Salmelin-errors})].
As a result the comparison shows inconsistencies between the size of the electron bubble as determined 
from the normal state mobility, $k_fR=11.17$, and the hard sphere radius that was used to account for
the longitudinal mobility in the superfluid phase, $k_f R = 16$. Even with this much larger electron 
bubble radius, the calculated Hall ratio, $\eta_{xy}/\eta_{xx}$, based on the formulae of 
Refs.~[\onlinecite{sal89,sal90}], is a factor of two to four smaller than the observed
Hall effect.\cite{ike13}

In the following we show that the scattering theory for the Stokes tensor for electron bubbles 
moving in \Hea\ presented in Secs.~(\ref{sec-Ion_Structure-3He-Normal}) - (\ref{sec-Scattering_Theory})
provides a quantitative account of the magnitude and temperature dependences of both the longitudinal 
mobility and the anomalous Hall effect within the hard-sphere model for the interaction 
of \He\ quasiparticles with the electron bubble. The only parameter in the theory is the hard sphere radius 
which we determine by fitting the transport cross-section for hard-sphere scattering to the normal-state 
mobility to obtain $k_f R = 11.17$. The electron-quasiparticle interaction is then determined by the 
hard sphere scattering phase shifts in Eq.~(\ref{delta_l_hs}), and plotted in 
Fig.~(\ref{figure_hard-sphere_phase-shifts}).

The calculations presented here for the transport cross sections and resultant components of 
the Stokes tensor are obtained by first solving the linear integral Eqs.~(\ref{Eq_1})-(\ref{Eq_4}) for 
the $t$-matrix amplitudes, $t^{m}_{a}(u',u)$. We transform the integral equations to coupled algebraic 
equations using Gauss-Legendre quadrature rules of even order, and integrate the square-root singularities 
appearing in the propagators following the procedure given in Ref.~[\onlinecite{Abramowitz72}].
  
\subsection{Scattering Cross Sections}

\begin{figure}[!]
\centering
\includegraphics[width=0.375\textwidth,keepaspectratio]{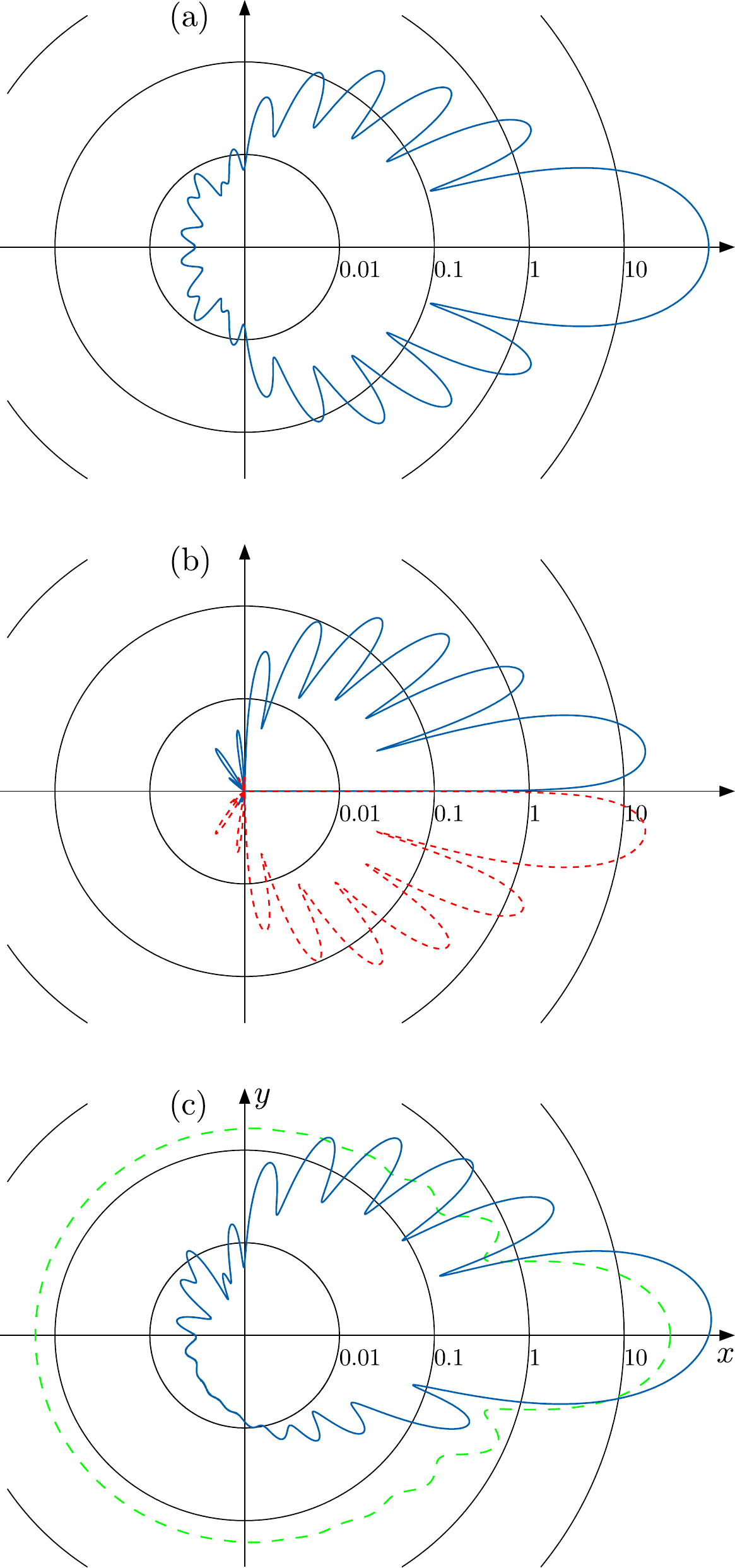}
\caption{(Color online) 
Panel (a) shows a polar plot of the differential cross 
section, $d\sigma^{(+)}(\hat\vk',\hat\vk;E)/d\Omega_{\vk'}$, 
as a function of the in-plane scattering angle angle $\phi'-\phi$ 
[Eqs.~(\ref{W_+-}) and (\ref{dsigma-dOmega})],
with incoming and outgoing momenta lying in the $xy$-plane, i.e. $\theta'=\theta=\pi/2$ and 
$\hat\vk=\ve_x$ ($\phi=0$).
The contours mark the magnitudes of the differential cross-sections in units of $\pi R^2$ on a log-scale.
The quasiparticle energy is $E = 1.01\Delta$, and the ion-quasiparticle potential is a 
hard sphere with $k_f R = 11.17$.
Similarly, panel (b) shows the asymmetry in the angular distribution of scattered 
quasiparticles given by $d\sigma^{(-)}(\hat\vk',\hat\vk;E)/d\Omega_{\vk'}$, which changes 
sign continuously across the lines $\Delta\phi=0,\pi$. The sign change is indicated by the 
dashed red curve.
Panel (c) shows the sum of these two differentical cross-sections, highlighting the 
asymmetry in the angular distribution of scattering quasiparticles for $\vl || \vz$.
The angular distribution for quasiparticle-ion scattering in the normal state is shown 
as the dashed green line. 
}
\label{fig-differential_cross-section}
\end{figure}

\begin{figure}[t]
\centering
\includegraphics[width=0.50\textwidth,keepaspectratio]{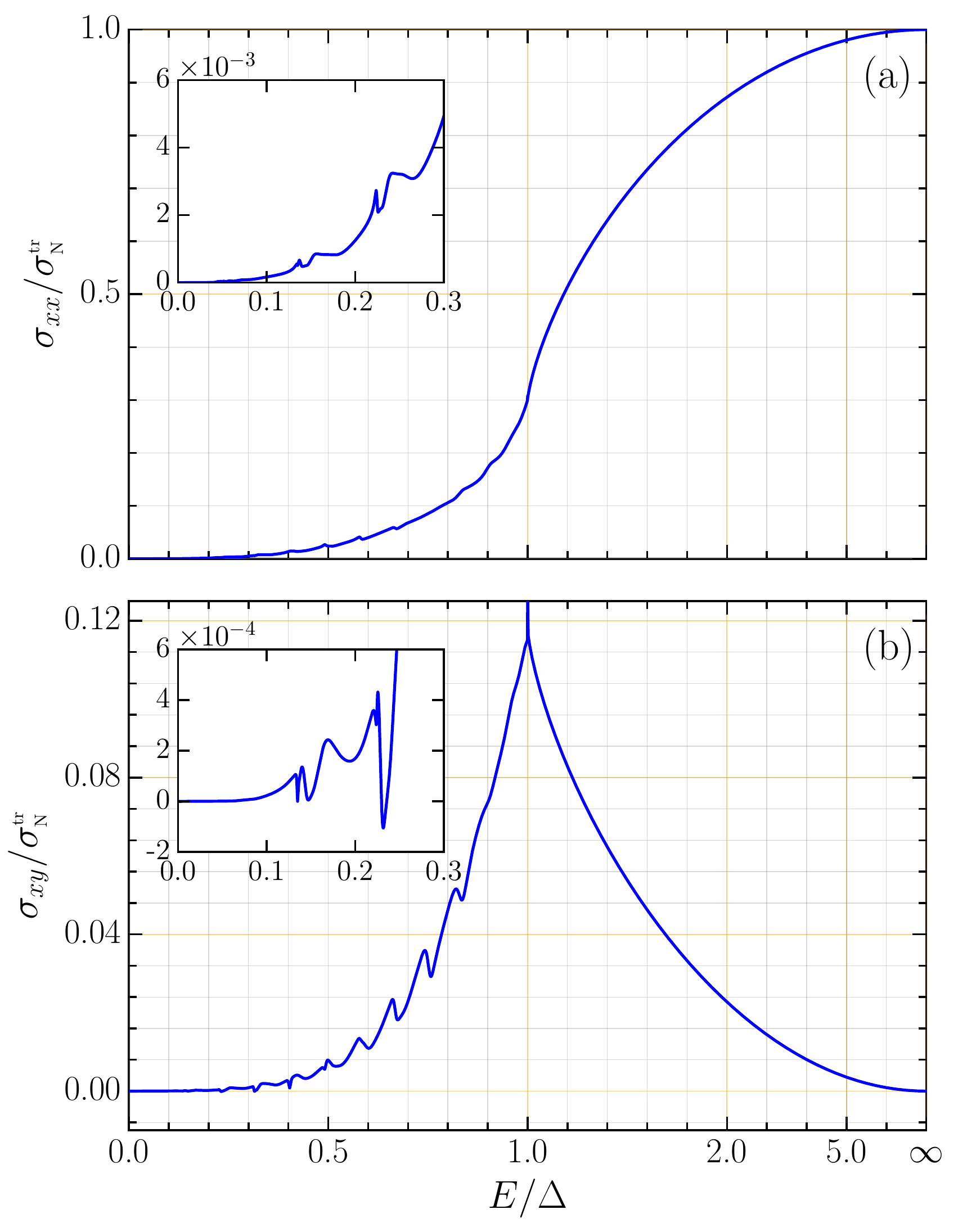}
\caption{The longitudinal [panel (a)] and transverse [panel (b)] transport cross sections as a function 
of energy for $T = 0.5T_c$.
The peak-dip structure at energies below the maximum gap, $\Delta$, are resonances originating from 
scattering  of quasiparticles by chiral Fermions bound to the surface of the electron bubble associated 
with distinct angular momentum channels [Eqs.~(\ref{sigma_+})-(\ref{sigma_-})]. 
The quasiparticle-ion potential is a hard sphere with $k_f R = 11.17$. The insets highlight the low-energy 
region.}
\label{Fig5}
\end{figure}

In Fig.~(\ref{fig-differential_cross-section}) we show results for the differential cross section 
defined in Eqs.~(\ref{W_+-})
and (\ref{dsigma-dOmega}) for in-plane scattering, i.e. both incident, $\vk$, and scattered, $\vk'$, 
wavevectors in the $xy$-plane. 
In particular, for an incoming quasiparticle with $\hat{\vk} = \ve_x$ ($\theta=\pi/2,\phi=0$) 
the symmetric part of the angular distribution, $d\sigma^{(+)}/d\Omega_{\vk'}$, contributing to 
$\sigma^{(+)}_{ij}(E)$ is shown in panel (a), and the asymmetry in the angular distribution, 
$d\sigma^{(-)}/d\Omega_{\vk'}$, of the scattered excitations is shown in panel (b)
as a function of the azimuthal scattering 
angle $\Delta\phi=\phi'-\phi$. Note that $d\sigma^{(-)}/d\Omega_{\vk'}$
changes sign across the lines $\Delta\phi=0$ and $\Delta\phi=\pi$,
and determines the anti-symmetric, transverse cross section, $\sigma^{(-)}_{ij}(E)$.
%
%
The total differential cross-section is shown in Fig.~(\ref{fig-differential_cross-section}c) 
in comparison with that for quasiparticle-ion 
scattering in the normal state. There is strong reduction in backscattering in the superfluid state compared
to that in the normal state, as well as the sharp angular dependences associated with resonant scattering 
from the spectrum of chiral Fermions bound to the ion, evident in the angular momentum resolved density of 
states shown in Fig.~(\ref{Fig_LDOS_res_m}). 
Resonant scattering of quasiparticles by the spectrum of chiral Fermions bound to the electron bubble is 
also evident in the energy-resolved transport cross-sections, $\sigma^{(-)}_{xy}(E)$ and $\sigma^{(+)}_{xx}(E)$,
shown in Fig~(\ref{Fig5}) normalized by the normal state transport cross section.

One can clearly see the peak-dip structure at energies below the maximum gap $\Delta$. These structures are
due to resonant scattering from chiral Fermions bound to the surface of the electron bubble. There is a 
resonance for each angular momentum channel $m$.
The chiral Fermions form as a result of multiple potential and Andreev scattering of quasiparticles off the 
electron bubble and the chiral order parameter in which it is embedded. This multiple scattering and bound 
state formation is encoded in the $t$-matrix equations of Eqs.~(\ref{Eq_1}) - (\ref{Eq_4}).

\subsection{Forces on moving electron bubbles}

\begin{figure}[t]
\centering
\includegraphics[width=0.49\textwidth,keepaspectratio]{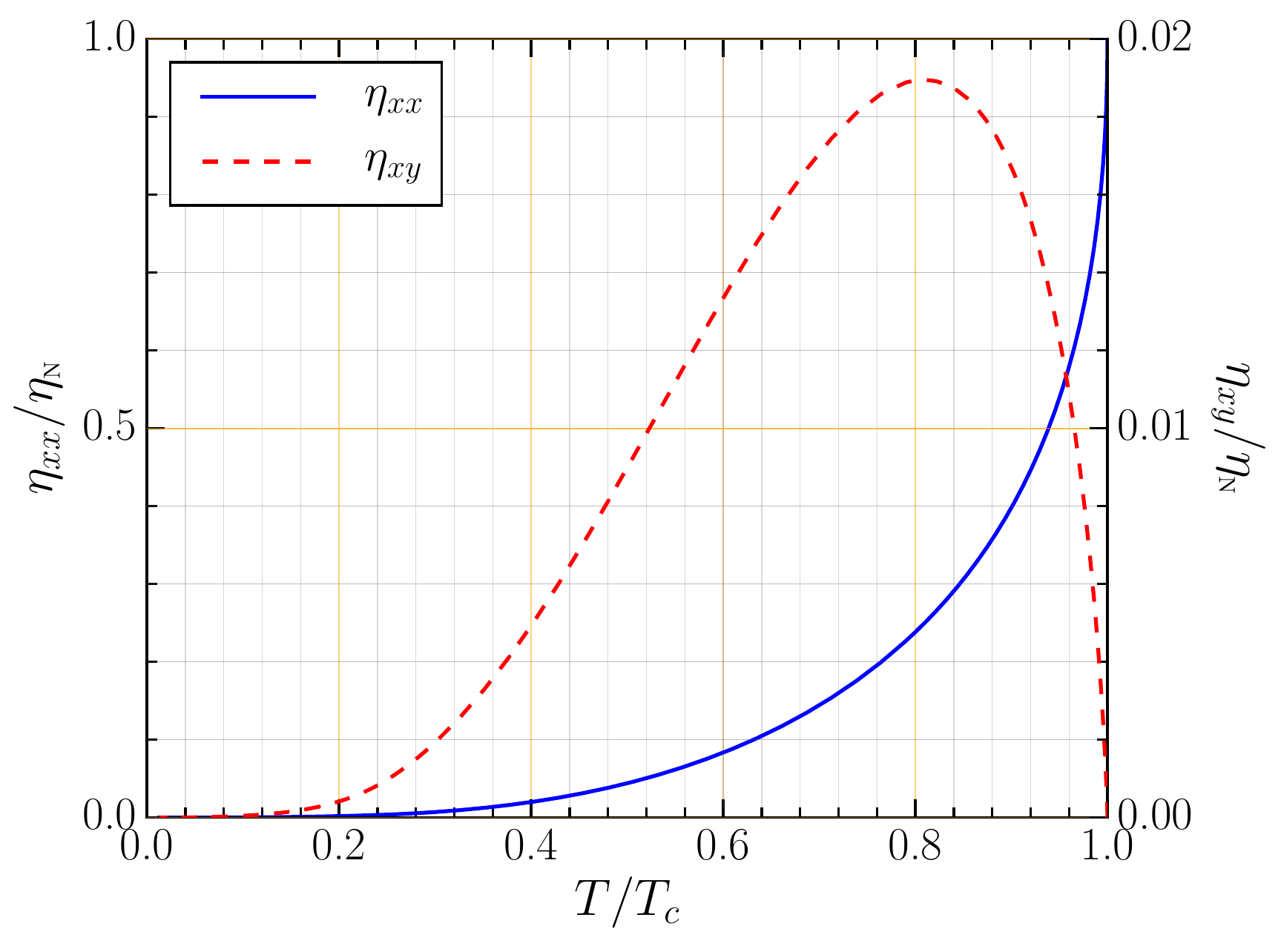}
\caption{(Color online) Longitudinal and transverse Stokes parameters, 
$\eta_{xx}/\eta_{\text{N}}$ (solid blue line) and $\eta_{xy}/\eta_{\text{N}}$ (dashed red line), 
as a function of $T/T_c$. Calculations are based on the hard sphere 
quasiparticle-ion potential with $k_f R=11.17$.
}
\label{Fig2}
\end{figure}

The transport cross sections, $\sigma^{(+)}_{ij}(E)$ and $\sigma^{(-)}_{ij}(E)$ calculated for the 
hard sphere potential, are used to calculate the components of the Stokes tensor given in 
Eq.~(\ref{eq-Stokes_Tensor}). 
In Fig.~(\ref{Fig2}) we show our results for the temperature dependences of the longitudinal  
($\eta_{xx}/\eta_{\text{N}}$) and transverse ($\eta_{xy}/\eta_{\text{N}}$) forces normalized to the 
normal state Stokes drag $\eta_{\text{N}}$.
The longitudinal drag force drops rapidly below $T_c$ due to the (i) opening of the gap in the bulk 
excitation spectrum and (ii) resonant scattering reflected in terms of strong suppression of backscattering
as shown in Fig.~(\ref{fig-differential_cross-section}).
The transverse force onsets at $T_c$, increases rapidly then decays at very low temperatures.

In the GL limit, $\Delta(T)/k_{\text{B}}T_c\sim(1-T/T_c)^{\nicefrac{1}{2}}\ll 1$, the drag force decreases as
$\eta_{xx}/\eta_{\text{N}}-1 \propto -\Delta(T)$, while the transverse force scales as
$\eta_{xy}/\eta_{\text{N}}\propto \Delta(T)^2\sim(1-T/T_c)$, reflecting the onset of branch conversion
scattering of Bogoliubov quasiparticles.
The scaling near $T_c$ follows from the GL expansion of the cross-sections given in App.~(\ref{App_CSs}). 
The scaling of $\eta_{xy}\sim\Delta(T)^2$ agrees with that inferred from the estimate given in Eq.~[1] of 
Ref.~(\onlinecite{sal89}); however, these authors include an additional small factor, $k_{\text{B}}T_c/E_f\sim 10^{-3}$, 
typically associated with normal-state particle-hole asymmetry. 
In our theory, particle-hole asymmetry is \emph{generated} by branch conversion scattering and particle-hole 
coherence that onsets at $T_c$, and is reflected in the asymmetric chiral spectrum for 
$d\sigma^{(-)}/d\Omega_{\vk'}(\vk',\vk;E)$. There is no factor, $k_{\text{B}}T_c/E_f$; however, there
is a small factor originating from the small \emph{transverse} momentum transfer that is a reflection 
of branch conversion scattering from the chiral order parameter. 
Our estimate of the longitudinal and transverse forces near $T_c$ for an electron bubble with 
velocity $v\,\ve_x$ is as follows.
For the moving ion encountering a flux $n\,v$, the typical momentum transfer imparted to the ion per 
quasiparticle (QP) 
collision is $\sim p_f$, and the momentum transport cross-section near $T_c$ is 
$\langle\sigma_{xx}\rangle\approx\sigma_N^{\text{tr}}\approx \pi R^2$, 
giving a drag force $|F_x| \approx n\,v\,p_f\,\sigma_N^{\text{tr}}$.
Now for branch conversion scattering there is angular momentum transfer of $\hbar$ by the chiral order 
parameter per branch conversion scattering of a QP. Thus, the transverse momentum transfer is of 
order $\hbar/R$ per QP.
Note that Andreev scattering is via the order parameter; there is no hard scattering with 
momentum transfer of order $p_f$. The fact that there is any momentum transfer is 
because of the angular momentum transfer via the chiral order parameter.  
In addition, branch conversion scattering onsets at $T_c$, thus the cross-section
is reduced relative to that for the longitudinal force by the probability of branch conversion 
scattering of thermal Bogoliubov QPs near $T_c$, i.e. 
$\langle \sigma_{xy}\rangle \approx (\Delta(T)/k_{\text{B}}T_c)^2\,\sigma_{N}^{\text{tr}}$, 
leading to 
$|F_y| \approx n\,v\,(\hbar/R)\langle\sigma_{xy}\rangle\approx 
n\,v\,(\hbar/R)\,\sigma_N^{\text{tr}}(\Delta(T)/k_{\text{B}}T_c)^2$,
and the ratio\footnote{{The force ratio estimate given in Eq.~(\ref{eq-force_ratio-GL}) was obtained by 
Vladimir Mineev based on hydrodynamic scaling in the Knudsen and GL limits (private communication).
Our analysis gives the same result, and is based on our scattering theory formulation for potential 
scattering and branch conversion scattering.}}
\be\label{eq-force_ratio-GL}
\frac{|F_y|}{|F_x|}\simeq\frac{1}{k_f R}\,\left(\frac{\Delta(T)}{k_{\text{B}}T_c}\right)^2
\,.
\ee 
The factor $1/k_f R$ accounts for the relative size of the transverse
and longitudinal transport cross-sections at $E\approx \Delta $ shown in
Fig.~(\ref{Fig5}), and also accounts for the order of magnitude reduction in the 
ratio $\eta_{xy}/\eta_{xx}$ shown in Fig.~(\ref{Fig2}) at $T/T_c\approx 0.8$. 
Note that the transport cross sections, $\sigma_{xx}(E)$ and $\sigma_{xy}(E)$, were both defined 
by scaling out the dimensional factors of $p_f$ in the kinematics. Thus,
$\sigma_{xy}(E)\simeq\nicefrac{\hbar}{p_f R}\sigma_{xx}(E)$ at $E\approx\Delta$. The
spectral average, $\langle\sigma_{xy}(E)\rangle$ near $T_c$ generates the additional 
factor of $(\Delta/k_{\text{B}}T_c)^2$.
Although the transverse force is roughly an order of magnitude smaller than the drag force, 
it leads to a dramatic effect on the dynamics of the negative ion.

The equation of motion for an electron bubble under the action of an in-plane electric field is
\be\label{eq-bubble_dynamics}
M\der{\vv}{t} = e\vE - \eta_{\perp}\,\vv -\eta_{xy}\,\vv\times\hat\vl  
\,,
\ee
where $M$ is the effective mass of the electron bubble. The first term on the right side of 
Eq.~(\ref{eq-bubble_dynamics}) is the Coulomb force on the ion, the second term is the drag force 
on the moving electron bubble, and the third term is the transverse force from the scattering of 
quasiparticles off Weyl Fermions bound to the ion.
The drag force results in relaxation of the ion velocity on a timescale $\tau$ given by 
$1/\tau = \eta_{\perp}/M$, while the transverse force has the form of the Lorentz
force, $\vF_{\text{W}} = \nicefrac{e}{c}\,\vv\times\vB_{\text{W}}$, where the effective 
magnetic field arises from scattering of quasiparticles off the Weyl spectrum of the ion,
\ber
\vB_{\text{W}} &=& -\frac{c}{e}\eta_{xy}\,\hat\vl
\nonumber
\\
&\approx& \frac{\Phi_0}{3\pi^2}\,k_f^2\,(k_f R)^2\,\left(\frac{\eta_{xy}}{\eta_{\text{N}}}\right)\,\hat\vl
\,,
\eer
where $\Phi_0=hc/2|e|$ is the flux quantum and we have approximated the normal state transport 
cross section by $\sigma_{\text{N}}^{\text{tr}}\approx \pi R^2$. Note that the temperature dependence
of $B_{\text{W}}$ is shown in Fig.~(\ref{Fig2}), and thus the order of magnitude of the Weyl field ranges 
from $B_{\text{W}} = 10^4\,\mbox{T}$ at $T/T_c = 0.8$ to $B_{\text{W}} = 10^3\,\mbox{T}$ at $T/T_c = 0.3$,
orders of magnitude larger that any laboratory magnetic field.\cite{ike15}

The Weyl field and drag force generate damped cyclotron motion of the electron bubble with frequency,
$\omega_{\text{c}} = eB_{\text{W}}/Mc$. The resulting steady-state velocity of the electron bubble in the 
combined electric ($\vE = \cE\ve_x$) and Weyl ($\vB_{\text{W}}=B_{\text{W}}\ve_z$) fields is given by
\ber
v_x = \frac{\tau}{1+(\omega_{\text{c}}\tau)^2}\,e\cE  
\,,
\quad
v_y = \frac{\tau(\omega_{\text{c}}\tau)}{1+(\omega_{\text{c}}\tau)^2}\,e\cE 
\,.
\eer
The transverse component is the anomalous Hall current, and the ratio with the longitudinal current
gives the Hall angle,
\be\label{eq-Hall_ratio}
\tan\alpha = \frac{v_y}{v_x} 
                      = \omega_{\text{c}}\tau 
                      = \frac{eB_{\text{W}}}{Mc}\tau = \frac{\eta_{xy}}{\eta_{\perp}}
\,.
\ee
Note that in spite of the enormous effective magnetic field, the Hall angle is relatively small because the 
relaxation time $\tau$ is so short compared to the cyclotron period, i.e. the drag force dominates the 
transverse force.
At $T/T_c =0.8$, where the Weyl field is maximum, the Hall angle is of order 
$\tan\alpha = \eta_{xy}/\eta_{\perp} \approx 0.1$. The detailed temperature dependences of 
the Stokes parameters show that the maximum Hall angle is $\tan\alpha_{\text{max}} \approx 0.25$
at $T/T_c \approx 0.4$, as shown in Fig.~(\ref{Fig4}), and discussed in more detail in comparison with
the experimental measurements below.

\subsection{Comparison between Theory and Experiment}\label{sec_Experiment-Theory}

\begin{figure}[t]
\centering
\includegraphics[width=0.475\textwidth,keepaspectratio]{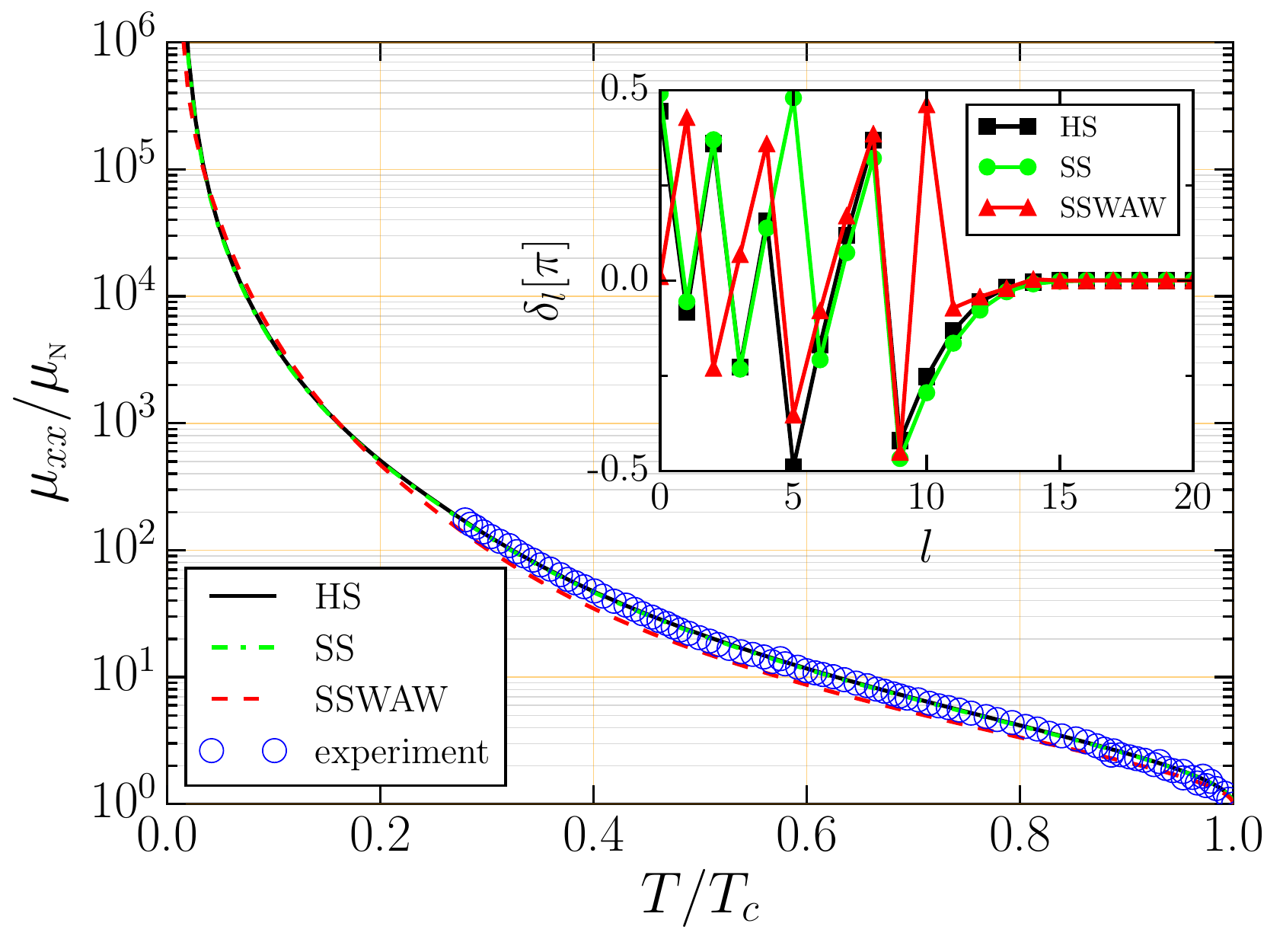}
\caption{
Experimental data for the longitudinal mobility normalized to the normal-state mobility is from 
Ref.~[\onlinecite{ike13}] shown as blue circles.
The theoretical result based on the hard sphere (HS) quasiparticle-ion potential is the black curve.
Results based on the soft core (SS) potential are shown as the dashed green curve, and those
for the four-parameter potential with intermediate attraction (SSWAW) are shown as the red dashed curve.  
Inset: Scattering phase shifts vs. angular momentum channel calculated for the three potentials. For
the HS model: $k_f R=11.17$; SS model: $V_0=1.01\,E_f$ and $k_f R = 12.48$; SSWAW model: $V_0=100\,E_f$
$V_1=10\,E_f$, $k_fR'=10.99$ and $R/R'=0.36$, all constrained by 
the experimental value of $\mu_{\text{N}}$.}
\label{Fig1}
\end{figure}

The experimental results for the transport of electron bubbles in \He\ are presented in terms of 
the components of the mobility tensor. The components of the mobility tensor are calculated from 
the Stokes parameters using Eqs.~(\ref{mobility-Stokes}). 
In Fig.~(\ref{Fig1}) we compare our theoretical result for the longitudinal mobility based on numerical 
calculations, using the machinery presented in the previous sections, with the experimental data reported 
in Refs.~[\onlinecite{ike13,ike15}]. 
The hard sphere potential works remarkably well, reproducing the longitudinal mobility data for \Hea\ over 
nearly two and a half decades for $0.25 \lesssim T/T_c \le 1$. 
It is worth emphasizing that the hard sphere potential is a single-parameter potential with the radius, 
$k_f R = 11.17$, fixed by the normal-state mobility. There are no other adjustable parameters in the 
theory, thus the comparison between theory and experiment for $\mu_{xx}/\mu_{\text{N}}$ is essentially 
perfect down to $T\approx 250\,\mu\mbox{K}$.

\begin{figure}[t]
\centering
\includegraphics[width=0.45\textwidth,keepaspectratio]{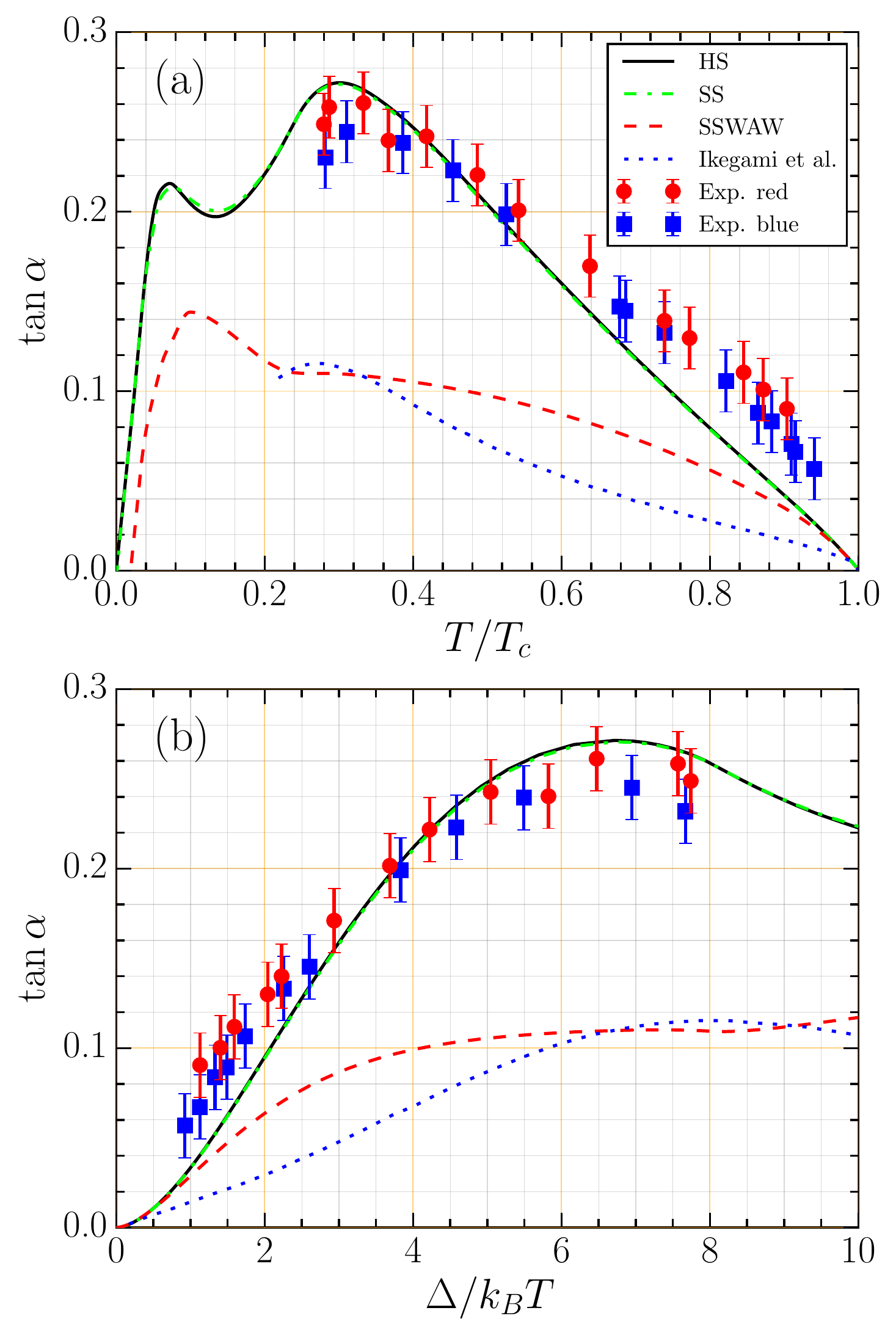}
\caption{(Color online) Panel (a): Hall ratio for the motion of electron bubbles in \Hea, 
$\tan\alpha \equiv v_y/v_x = \eta_{xy}/\eta_{xx}$, as a function of temperature 
for the hard sphere model for the quasiparticle-ion potential with $k_f R=11.17$ 
(black line).
The experimental data was from the RIKEN group.\cite{ike13,ike15} 
Theoretical results for the repulsive soft core potential (dashed green line) and 
repulsive potential with short-range attraction (dashed red line) are
shown for comparison.
The dotted blue line corresponds to the calculation based on  
the formulas from Salmelin \textit{et al.}\cite{sal89,sal90} presented in 
Refs~[\onlinecite{ike13,ike15}]. 
Panel (b): The same results presented as a function of $\Delta(T)/k_BT$.
}
\label{Fig4}
\end{figure}

We note that Ikegami et al.\cite{ike13} report a reasonably good comparison with their data, albeit 
with observable deviations 
at lower temperatures, using the incorrect formula for $\mu_{xx}/\mu_{\text{N}}$ from 
Ref.~[\onlinecite{sal90}] with a hard sphere radius of $k_f R=16$. 
This much larger value disagrees with the radius obtained from measurements of the normal-state mobility.
Moreover, as the authors of Ref.~[\onlinecite{ike13}] found, the formula for the transverse mobility, 
$\mu_{xy}$, from Ref.~[\onlinecite{sal90}] is in serious disagreement with experimental measurements of 
the transverse mobility as it under estimates the Hall angle by a factor of $\approx 2 - 4$ over a 
large temperature range, $0.25T_c \lesssim T \le T_c$, based on the same value of $k_f R$. Again, the 
discrepancy originates from an incorrect formula for $\mu_{xy}$ reported in 
Ref.~[\onlinecite{sal90}] (see App.~(\ref{appendix_Salmelin-errors})).\footnote{The actual discrepancy 
is more severe. The theory of Salmelin et al. in 
Refs. [\onlinecite{sal89,sal90}], when evaluated properly, predicts 
\emph{zero transverse force}, i.e. $\mu_{xy}\equiv 0$. The expression used for calculating 
$\mu^{-1}_{xy}$ by the RIKEN group, Eq. [6] and Eq. [11] from Ref.~[\onlinecite{sal90}], is 
\emph{identically zero} when evaluated with the correct angular dependence for the kinematic factor, 
$(\hat{\vk}'-\hat{\vk})_x(\hat{\vk}'-\hat{\vk})_y$.}

While the comparison of our theoretical prediction for $\mu_{xx}$ is excellent agreement with the 
RIKEN measurements, the strong test is the comparison of our calculations for the transverse 
force with the measurements of the anomalous Hall effect.
In Fig.~(\ref{Fig4}) we 
show our theoretical results [solid (black) curves] for the anomalous Hall ratio given by 
Eq.~(\ref{eq-Hall_ratio}), with the calculated results for $\eta_{xy}$ and $\eta_{xx}$ (shown in 
Fig.~(\ref{Fig2})), plotted vs. $T/T_c$ in panel (a), and vs. $\Delta(T)/k_{\text{B}}T$ 
in panel (b).
The (red) circular [(blue) square] symbols correspond to the experimental data reported in 
Refs.~[\onlinecite{ike13,ike15}].
For comparison we include the results of the calculation by Ikegami et al. based on the formulae from 
Ref.~[\onlinecite{sal90}] as the dotted (blue) lines. 

It seems worth re-emphasizing that in all our calculations reported here the only parameter is 
hard sphere radius for the quasiparticle-ion potential which is fixed at the outset as 
$k_f R = 11.17$ by the normal-state mobility. 
Thus, we view the overall agreement between theory and experiment as strong confirmation
of the scattering theory, particularly the origin of the anomalous Hall effect resulting
from resonant scattering of thermal quasiparticles by the spectrum of Weyl fermions bound to the
electron bubble embedded in \Hea.

The theoretical prediction shown in Fig.~(\ref{Fig4}) shows structure in the 
Hall ratio - a dip-peak structure  - below $T\approx 0.25\,T_c$.
An important test of this theory would be measurements of the Hall mobility extended 
below $0.2\,\mbox{mK}$.


\subsection{Beyond the hard sphere potential}

Although the hard sphere model for the quasiparticle-ion potential provides very good agreement 
with the observed forces acting on the moving ion, it is a only rough approximation to 
expectations of the microscopic interaction between \He\ quasiparticles and the electron
bubble. To test the robustness of our theoretical predictions to the quasiparticle-ion potential
we consider a more general central potential with short-range repulsion and intermediate-range 
attraction, 
\be
V(r) = 
\begin{cases}
V_0,  &r \leq R
\,,\\
-V_1, &R < r \leq R'
\,,\\
0,    &r > R'
\,.
\end{cases}
\label{soft_sphere_waw}
\ee
The normal-state scattering phase shifts for this piece-wise constant potential are expressed in terms of regular and 
modified spherical Bessel functions; the analytical formulas are given in 
Eqs.~(\ref{eq-deltas-four-parameter_potential})-(\ref{eq-deltas-betas}) 
of App.~\ref{appendix_phase-shifts}.
We discuss two cases both with $V_0>E_f$: (i) for $V_1=0$ the potential is a two-parameter, repulsive ``soft-core'' potential,
and (ii) for $V_1>0$ and $R'>R$ we include in addition to the short-range repulsion, an intermediate range attraction.
The latter case allows for a shallow bound state, and corresponding scattering resonance, in one or more angular momentum
channels, $l\le l_{\text{max}}$.

Figures~\ref{Fig1} and \ref{Fig4} show our calculations for the longitudinal mobility and Hall ratio for these potentials
in comparison with the results for the hard sphere potential. The corresponding phase shifts are shown in the inset.
For the ``soft-core'' model we chose a weakly repulsive potential, $V_0 = 1.01\,E_f\approx 0.5\mbox{meV}$, and adjusted 
the radius $R$ to fit the measured normal-state mobility, 
$\mu^{\text{exp}}_{\text{N}}=1.7\times 10^{-6}\,\mbox{m}^2/\mbox{V}/\mbox{s}$, as was done for the hard-sphere potential.
The resulting phase shifts, shown in inset of Fig.~(\ref{Fig1}), are similiar to those of calculated for hard-sphere scattering 
in that there are no additional strong scattering channels; the phase shift for the $l=5$ channel corresponds to strong scattering 
for both the hard sphere and the soft core potential. 
Furthermore, there is virtually no observable change in the theoretical predictions for the longitudinal and transverse forces 
on the moving ion described by the soft core potential, compared to the results for the hard sphere potential.  
This is representative of the general class of short-range repulsive potentials. So long as the 
range of the repulsive quasiparticle-ion potential is adjusted the fit the normal state mobility we obtain excellent agreement
for the forces on the negative ion in the superfluid phase.\cite{she16b}

The situation is different for the case with short-range repulsion and intermediate range attraction. 
Here we fixed $V_0= 100\,E_f$ and $V_1 = 10\,E_f$, then adjusted $R$ and $R'$ to obtain a best fit to the experimental 
value of the normal-state mobility, giving $k_fR' = 10.99$ and $R/R'=0.36$.
As can be seen from the inset of Fig.~\ref{Fig1}, the intermediate range attraction changes the set of scattering phase shifts, 
compared to the hard sphere potential, with the most dramatic change happening for $l=10$. This channel exhibits an additional
scattering resonance [red triangles in the inset of Fig.~(\ref{Fig1})].
The scattering of quasiparticles in this channel is enhanced towards the unitary limit, $\delta_{l=10} \approx \pi/2$, which makes 
the partial scattering cross section for this channel maximal. 
As a consequence, the forces on the ion are modified. The longitudinal mobility shown in Fig.~\ref{Fig1} 
(red dashed line) is slightly reduced compared to that for the hard sphere scattering potential.
More dramatic is the reduction in the anomalous Hall ratio shown in Fig.~(\ref{Fig4}), which deviates strongly from the 
experimental data (red dashed line). The main conclusion here is that for the negative ion the quaiparticle-ion scattering
potential is repulsive and short range, and the experimental results are well described by hard sphere potential scattering.

A softer core potential with intermediate range attraction may be relevant to understanding the mobility of positive ions in 
\Hea, given that the positive ion attracts \He\ to form a ``snowball'' of \He\ atoms with increased density relative to 
bulk \He.\cite{dobbs00} Indeed preliminary measurements of the longitudinal and transverse forces on a positive ion in \Hea\ show 
different magnitudes and temperature dependences for the longitudinal mobility and anomalous Hall ratio compared to 
the negative ion.\cite{ike15} However, a detailed theoretical description of the structure and transport properties of the positive ion is 
outside the scope of this report.  

\section{Discussion}

The comparison between theory and experiment for the Hall ratio shows a maximum deviation 
of  $\approx 15\%$ at $T\approx 0.8T_c$, which is the temperature at which the transverse 
force, $\eta_{xy}$, is a maximum. This suggests that there may be an additional contribution 
to the transverse force on the moving ion.
Within the theory of thermal quasiparticles scattering off the moving ion, the larger experimental
value for $\eta_{xy}$ suggests an additional weak scattering mechanism contributing to
the transport cross section, $\sigma_{xy}(E)$, at energies close to the gap edge, or perhaps
deviations from the hard sphere potential. 
These possibilities for an additional contribution to the transverse force on the moving 
ion are addressed in a separate report.

It is also likely that in the low temperature limit, $T <0.25 T_c$, 
new physics appears in the transport of electron bubbles in \Hea.
In particular, the theoretical prediction of the sub-gap spectrum shown in 
Fig.~(\ref{Fig1}) leads to the sharp increase in the longitudinal mobility 
at low temperatures.
Thus, at constant electric field we expect the linear theory for the Stokes force tensor
to fail at sufficiently low temperatures as there is insufficient drag force from
thermal quasiparticles to limit the ion velocity below the Landau critical velocity, 
$v_c = \Delta/p_f$. At high velocity the ion will dissipate energy by Cherenkov radiation 
of quasiparticles.\cite{jen88} 
This process may onset at velocities well below $v_c$ given the low energy Weyl 
spectrum near the moving ion, and it is an open question as to whether and how the resulting 
quasiparticle radiation might contribute to transverse force.

\vspace*{-5mm}
\begin{acknowledgments}

The research of OS and JAS was supported by the National Science Foundation (Grant DMR-1508730).
We acknowledge key discussions with Hiroki Ikegami, Kimitoshi Kono and Yasumasa Tsutsumi on the
RIKEN electron mobility experiments that provided the motivation for this study.
We thank Vladimir Mineev for discussions on the magnitude and interpretation of the origin of 
the transverse force.
\end{acknowledgments}
\appendix

\vspace*{-3mm}
\section{Critique of Salmelin and Salomaa's Theory}\label{appendix_Salmelin-errors}

The report by Salmelin and Salomaa (SS) on the mobility of electron bubbles in superfluid \Hea\ 
was an attempt to extend the earlier work by Salomaa et al.\cite{sal80} on the same topic 
to calculate the transverse component of the mobility, $\mu_{xy}$. The latter was argued 
in Ref.~[\onlinecite{sal89}] to exist based on the analogy of the Magnus effect for a 
spinning object moving through a fluid, in this case the electron bubble with bound 
circulating currents.
While the physical argument in Ref.~[\onlinecite{sal89}] for the transverse component 
of the mobility is sound, the formulation of the scattering theory by 
Salmelin et al.\cite{sal89,sal90} 
cannot account for the transverse force on a moving electron bubble. 

The primary error introduced by Salmelin et al.\cite{sal89,sal90} in their formulation 
of the transport cross section for an electron bubble moving in superfluid \Hea\ is the 
assumption of microscopic reversibility for scattering rates for the transition 
$\vk\rightarrow\vk'$ and the inverse scattering event, $\vk'\rightarrow\vk$, i.e. that 
$W(\vk',\vk)=W(\vk,\vk')$. 
However, \Hea\ breaks mirror symmetry in any plane containing the chiral axis $\hat\vl$, as well 
as time-reversal symmetry. Thus, the condition on the scattering rate for quasiparticles scattering off an 
ion in \Hea\ connects the two scattering events for mirror reflected ground states, 
i.e. $W(\vk',\hat\vk;+\hat\vl) = W(\vk,\vk';-\hat\vl)$. Conversely, microscopic reversibility is violated for 
the broken symmetry ground state with fixed chirality $+\hat\vl$.

By assuming microscopic reversibility the authors of Ref.~[\onlinecite{sal90}] pre-supposed mirror symmetry 
in the scattering of quasiparticles off the electron bubble, and thus ensured that the Stokes tensor 
is symmetric and diagonal, i.e. that $\eta_{xy} = 0$. This conclusion is clear from Eqs.~[3],[5] and [6] of 
Salmelin et al.\cite{sal90}, and in the paragraph preceding Eqs. [4] of Ref.~[\onlinecite{sal89}].  
It is worth noting that the same assumption was made in the earlier work 
of Salomaa et al.\cite{sal80} for which there was no mention or calculation of a transverse force
on the moving ion.

So, why do SS obtain a non-zero result for the transverse mobility? 
They introduce a second error in the evaluation of the kinematic factors,
$(\hat\vk'-\hat\vk)_i(\hat\vk'-\vk)_j$ [$\Delta \vp_i\Delta \vp_j$ in the notation of SS]. Specifically,
Eqs.~[11] in SS are incorrect in their entirety. The argument in the paragraph preceding these formulae is
the source of the error. SS generated Eqs.~[11] by first assuming $\hat\vk=\hat\ve_x$ is fixed in the 
laboratory coordinate system such that the azimuthal angle $\phi_{\vk} = 0$. Then, the azimuthal angle
for the final state momentum, $\vk'$, was replaced by $\phi_{\vk'}\rightarrow \phi_{\vk'} - \phi_{\vk}$ 
to arrive at SS's Eqs.~[11]. This procedure is invalid, but has the effect of violating mirror symmetry
in the kinematics. All kinematic factors, $\Delta\vk_i\Delta\vk_j$, are invariant under the mirror operation
$\vk\leftrightarrow\vk'$, in particular, $(\hat\vk'-\hat\vk)_x(\hat\vk'-\hat\vk)_y$ is invariant under 
$\vk\leftrightarrow\vk'$, or equivalently under $\phi_{\vk}\leftrightarrow\phi_{\vk'}$. 
Eq.~[11] of SS for $\Delta\vk_x\Delta\vk_y$ violates mirror symmetry.

The result is a spurious transverse force from a mirror symmetric scattering rate. 
The violation of the mirror symmetry in the kinematic factors also predicts a spurious aniostropy 
of the drag force in the x-y plane, i.e. $\mu_{xx}\ne\mu_{yy}$, even in the isotropic normal 
Fermi liquid.
The authors recoginized the violation of the axial symmetry of A phase excitation gap, so they enforced a 
single in-plane drag coefficient by replacing 
$\Delta\hat\vk_x\Delta\hat\vk_x \rightarrow 
\nicefrac{1}{2}\left(\Delta\hat\vk_x\Delta\hat\vk_x + \Delta\hat\vk_y\Delta\hat\vk_y\right)$ 
in the calculation of $\mu_{\perp}^{-1}$.

The erroneous set of Eqs. [11] in SS for the momentum transfer factors invalidates all the calculations 
of cross sections and components of the mobility tensor in Ref.~[\onlinecite{sal90}] as well as Eqs.~[4] 
in Ref.~[\onlinecite{sal89}], and thus the source and magnitude of the transverse force on the moving
electron bubble. In particular, the theory of SS, when evaluated with the correct formulae for the 
kinematic factors, $\Delta\hat\vk_i\Delta\hat\vk_j$ yields only uniaxal Stokes drag forces and 
zero transverse force on the moving ion, as was originally obtained in Ref.~[\onlinecite{sal80}].

Our formulation of the force on the moving ion incorporates broken time reversal and mirror symmetries
by the \Hea\ ground state correctly. We are able to identify scattering events that contribute to 
the Stokes drag and the transverse force as, $W^{(+)}(\vk',\vk) = + W^{(+)}(\vk,\vk')$ and 
$W^{(-)}(\vk',\vk) = - W^{(-)}(\vk,\vk')$, respectively. 
Mirror symmetric scattering generates the drag forces, while the anti-symmetric component to the 
rate is responsible for the transverse force and the anomalous Hall effect, as we discuss in 
Sec.~(\ref{sec-cross-sections}).    

\onecolumngrid
\section{Kernel for the LDOS near the electron bubble}\label{appendix-LDOS_kernel}

The kernel, $K_{l'l}^m(u',u,\varepsilon)$ (Eq.~(\ref{LDOS_kernel}), defining the LDOS and the 
current density is obtained from the trace of the Nambu Green's function 
$\widehat\cG^{\text{R}}_{\text{S}}(\vr,\vr;E)$ in Eqs.~(\ref{eq-LDOS} - \ref{eq-Dyson_T-matrix}).
Only the $t$-matrix term in Eq.~(\ref{eq-Dyson_T-matrix}) contributes to the kernel, in which
case we are led to evaluate the integral  
\begin{equation}
\mathrm{I} = \int\frac{d^3k'}{(2\pi)^3}\int\frac{d^3k}{(2\pi)^3}
                                 e^{i(\mathbf{k}'-\mathbf{k})\cdot\mathbf{r}}
\widehat{G}^{\text{R}}_{\text{S}}(\vk',E)
\widehat{T}^{\text{R}}_{\text{S}}(\vk',\vk;E)
\widehat{G}^{\text{R}}_{\text{S}}(\vk,E)
\,.
\end{equation}
We use Eq.~(\ref{eq-G-Fourier}) and utilize the expansion of the plane wave,
$e^{i\mathbf{k}\cdot\mathbf{r}} = 4\pi\sum_{l=0}^{\infty}\sum_{m=-l}^{l}i^l
 j_l(kr)Y_l^m(\hat{\mathbf{k}})Y_l^m(\hat{\mathbf{r}})^{\ast}$,
in spherical harmonics and the regular spherical Bessel functions. 
In the quasiclassical limit, $E_f\ll\Delta$ (see Note~\onlinecite{Note5}), we evaluate the $t$-matrix 
in the elastic limit for momenta on Fermi surface and obtain
\begin{align}
\mathrm{I} &= (4\pi N_f)^2\sum_{l,l'=0}^{\infty}\sum_{m'=-l'}^{l'}\sum_{m=-l}^{l}i^{l'-l}
Y_{l'}^{m'}(\hat{\mathbf{r}})^{\ast}Y_l^m(\hat{\mathbf{r}})
\int\frac{d\Omega_{\mathbf{k}'}}{4\pi}\int\frac{d\Omega_{\mathbf{k}}}{4\pi}
Y_{l'}^{m'}(\hat{\mathbf{k}}')Y_l^m(\hat{\mathbf{k}})^{\ast}\notag\\
&\times
\left[\int_{-\infty}^{\infty}d\xi'j_{l'}(k'r)\widehat{G}^{\text{R}}_{\text{S}}(\vk',E)\right]
\widehat{T}^{\text{R}}_{\text{S}}(\hat\vk',\hat\vk;E)
\left[\int_{-\infty}^{\infty}d\xi j_{l}(kr)\widehat{G}^{\text{R}}_{\text{S}}(\vk,E)\right]
\,.
\label{integra;-I}
\end{align}
The remaining integral 
\begin{equation}
\mathrm{J} = \int_{-\infty}^{\infty}d\xi j_{l}(kr)\widehat{G}^{\text{R}}_{\text{S}}(\vk,E),
\end{equation}
is evaluated most conveniently using spherical Hankel functions of the first and second kind,
\begin{align}
h_l^{(1,2)}(x) = j_l(x) \pm in_l(x)
\,,
\quad\mbox{where}\quad
h_l^{(1,2)}(x)\propto e^{\pm ix}
\,,
\end{align}
in which case we obtain,
\begin{align}
&\mathrm{J} = \frac{1}{2}\left[h_l^{(1)}(k_fr)\,\mathrm{J}^{+} 
            + h_l^{(2)}(k_fr)\,\mathrm{J}^{-}\right]
\,,
\\
&\mathrm{J}^{\pm} \equiv \int d\xi \,e^{\pm i\frac{\xi}{\hbar v_f}r}\,
 \widehat{G}^{\text{R}}_{\text{S}}(\vk,E)
\,,
\end{align}
where we used $k = k_f + \xi/\hbar v_f$. The integrals $\mathrm{J}^{\pm}$ are evaluated 
using Eq.~(\ref{eq-GRS_bulk}),
\begin{equation}
\mathrm{J}^{\pm} = -i\pi e^{-\sqrt{|\Delta(\hat{\vk})|^2 - \varepsilon^2}\frac{r}{\hbar v_f}}
\left[
\frac{-i}{\sqrt{|\Delta(\hat{\vk})|^2 - \varepsilon^2}}
\begin{pmatrix}
\varepsilon\mathbb{1} & -\hat\Delta(\hat{\vk}) 
\\
-\hat\Delta^{\dagger}(\hat{\vk}) & \varepsilon\mathbb{1}
\end{pmatrix}
\pm
\begin{pmatrix} \mathbb{1} & 0\\ 0 & -\mathbb{1}\\ \end{pmatrix}
\right]
\,.
\end{equation}
Expressing the spherical harmonics as 
$Y_l^m(\hat{\mathbf{k}})\equiv Y_l^m(\theta,\phi) = \Theta_{l}^{m}(\cos\theta)\,e^{im\phi}$, 
we then integrate over the azimuthal angles in Eq.~(\ref{integra;-I}). Finally, Eq.~(\ref{LDOS_m}) 
is obtained by evaluating the trace over the Nambu matrices in Eq.~(\ref{integra;-I}),
\begin{equation}
\sum_{m=-\infty}^{\infty}\delta N_m(\mathbf{r},E) = -\frac{1}{2\pi}\mathrm{Im}\left[\mathrm{Tr}(\mathrm{I})\right].
\end{equation}

\section{Formulae for the electron bubble current density}\label{appendix-Current}

The current density circulating an electron bubble in cartesian components is 
$\vj(r,\vartheta,\varphi) = \sum_{i=x,y,z}\,j_{i}(\vr)\,\hat\ve_i$. The current along the chiral axis,
\ber
j_z(\mathbf{r}) 
&=& -4\pi^3v_fN_fk_BT\,\mathrm{Re}\Biggl\{
  \sum_{n=0}^{\infty}\sum_{m=-\infty}^{\infty}\sum_{l,l'=|m|}^{\infty}
  \Theta_{l'}^{m}(\cos\vartheta)\Theta_{l}^{m}(\cos\vartheta)\int_{-1}^{1}du'\int_{-1}^{1}du (u'+u)\notag\\
&\times&
  \Theta_{l'}^{m}(u')e^{-\sqrt{\Delta^2(1-u'^2)+\varepsilon_n^2}\frac{r}{\hbar v_f}}
  \Theta_{l}^{m}(u)e^{-\sqrt{\Delta^2(1-u^2)+\varepsilon_n^2}\frac{r}{\hbar v_f}} 
  K_{l'l}^m(u',u,i\epsilon_n)\Biggr\}
\,,
\eer

\noindent vanishes by symmetry; in particular, the spectrum of Weyl fermions is symmetric 
under $z\rightarrow -z$.
The in-plane components are expressed in terms for the four terms related to the components of the $t$-matrix, 
\begin{align}
&j_x(\mathbf{r}) = -4\pi^3v_fN_fk_BT\,\mathrm{Re}
\Biggl\{\sum_{n=0}^{\infty}\sum_{m=-\infty}^{\infty}\frac{1}{2}
\Bigl[
j_1^m(\mathbf{r},\epsilon_n) + j_2^m(\mathbf{r},\epsilon_n) 
+ 
j_3^m(\mathbf{r},\epsilon_n) + j_4^m(\mathbf{r},\epsilon_n)\Bigr]\Biggr\}
\,,\\
&j_y(\mathbf{r}) = -4\pi^3v_fN_fk_BT\,\mathrm{Re}\Biggl\{\sum_{n=0}^{\infty}\sum_{m=-\infty}^{\infty}\frac{1}{2i}
\Bigl[
j_1^m(\mathbf{r},\epsilon_n) - j_2^m(\mathbf{r},\epsilon_n) 
+ 
j_3^m(\mathbf{r},\epsilon_n) - j_4^m(\mathbf{r},\epsilon_n)
\Bigr]\Biggr\},
\end{align}
where
\begin{align}
j_1^m(\mathbf{r},\epsilon_n) 
&= 
e^{i\varphi}\sum_{l'=|m-1|}^{\infty}\sum_{l=|m|}^{\infty}
\Theta_{l'}^{m-1}(\cos\vartheta)\Theta_{l}^{m}(\cos\vartheta)\int_{-1}^{1}du'\int_{-1}^{1}du \sqrt{1-u'^2}
\notag\\
&\times
\Theta_{l'}^{m-1}(u')e^{-\sqrt{\Delta^2(1-u'^2)+\varepsilon_n^2}\frac{r}{\hbar v_f}}
\Theta_{l}^{m}(u)e^{-\sqrt{\Delta^2(1-u^2)+\varepsilon_n^2}\frac{r}{\hbar v_f}} K_{l'l}^m(u',u,i\epsilon_n)
\,,\\
j_2^m(\mathbf{r},\epsilon_n) 
&= e^{-i\varphi}\sum_{l'=|m+1|}^{\infty}\sum_{l=|m|}^{\infty}
\Theta_{l'}^{m+1}(\cos\vartheta)\Theta_{l}^{m}(\cos\vartheta)\int_{-1}^{1}du'\int_{-1}^{1}du \sqrt{1-u'^2}
\notag\\
&\times
\Theta_{l'}^{m+1}(u')e^{-\sqrt{\Delta^2(1-u'^2)+\varepsilon_n^2}\frac{r}{\hbar v_f}}
\Theta_{l}^{m}(u)e^{-\sqrt{\Delta^2(1-u^2)+\varepsilon_n^2}\frac{r}{\hbar v_f}} K_{l'l}^m(u',u,i\epsilon_n)
\,,\\
j_3^m(\mathbf{r},\epsilon_n) 
&= 
e^{i\varphi}\sum_{l'=|m|}^{\infty}\sum_{l=|m+1|}^{\infty}
\Theta_{l'}^{m}(\cos\vartheta)\Theta_{l}^{m+1}(\cos\vartheta)\int_{-1}^{1}du'\int_{-1}^{1}du \sqrt{1-u^2}
\notag\\
&\times
\Theta_{l'}^{m}(u')e^{-\sqrt{\Delta^2(1-u'^2)+\varepsilon_n^2}\frac{r}{\hbar v_f}}
\Theta_{l}^{m+1}(u)e^{-\sqrt{\Delta^2(1-u^2)+\varepsilon_n^2}\frac{r}{\hbar v_f}} K_{l'l}^m(u',u,i\epsilon_n)
\,,\\
j_4^m(\mathbf{r},\epsilon_n) 
&= 
e^{-i\varphi}\sum_{l'=|m|}^{\infty}\sum_{l=|m-1|}^{\infty}
\Theta_{l'}^{m}(\cos\vartheta)\Theta_{l}^{m-1}(\cos\vartheta)\int_{-1}^{1}du'\int_{-1}^{1}du \sqrt{1-u^2}
\notag\\
&\times
\Theta_{l'}^{m}(u')e^{-\sqrt{\Delta^2(1-u'^2)+\varepsilon_n^2}\frac{r}{\hbar v_f}}
\Theta_{l}^{m-1}(u)e^{-\sqrt{\Delta^2(1-u^2)+\varepsilon_n^2}\frac{r}{\hbar v_f}} K_{l'l}^m(u',u,i\epsilon_n).
\end{align}
Using the following symmetry properties of the kernel,
$K_{l'l}^m(u',u,i\epsilon_n) \equiv i^{l'-l}\kappa_{l'l}^m(u',u,i\epsilon_n)$,
\begin{align}
&\kappa_{l'l}^{-m}(u',u,i\epsilon_n) 
= 
-\left[\kappa_{l'l}^{m}(u',u,i\epsilon_n)\right]^{\ast}
\,,\\
&\kappa_{ll'}^{m}(u,u',i\epsilon_n) 
= 
\kappa_{l'l}^{m}(u',u,i\epsilon_n)
\,,
\end{align}
one finds that the current density is purely azimuthal, 
$\vj(r,\vartheta,\varphi) = j_{\varphi}(r,\vartheta)\,\ve_{\varphi} 
\,\,,\mbox{with}\,\,
\ve_{\varphi} = -\sin\varphi\ve_{x} + \cos\varphi\ve_{y}$,
\begin{align}
j_{\varphi}(r,\vartheta) 
&= 
-8\pi^3v_fN_fk_BT\,\sum_{n=0}^{\infty}
\sum_{m=-\infty}^{\infty}\sum_{l'=|m-1|}^{\infty}\sum_{l=|m|}^{\infty}
\Theta_{l'}^{m-1}(\cos\vartheta)\Theta_{l}^{m}(\cos\vartheta)\int_{-1}^{1}du'\int_{-1}^{1}du \sqrt{1-u'^2}
\notag\\
&\times
\Theta_{l'}^{m-1}(u')e^{-\sqrt{\Delta^2(1-u'^2)+\epsilon_n^2}\frac{r}{\hbar v_f}}
\Theta_{l}^{m}(u)e^{-\sqrt{\Delta^2(1-u^2)+\epsilon_n^2}\frac{r}{\hbar v_f}} 
\mathrm{Re}\left[\kappa_{l'l}^m(u',u,i\epsilon_n)\right]\mathrm{Im}
\left[i^{l'-l}\right]
\,.
\end{align}

\section{Formulae for the scattering rate and transport cross section}\label{App_CSs}

We summarize our results for the transport cross sections in terms of 
the solutions to the coupled Eqs.~(\ref{Eq_1})-(\ref{Eq_4}) for the $t$-matrix. 
Given the solutions for the branch components, $t_a^m$, we substitute 
Eqs.~(\ref{Eq_1})-(\ref{Eq_4}) into Eq.~(\ref{Scattering_Rate}), to obtain

\begin{align}
&W(\hat{\vk}',\hat{\vk})
= 
\frac{1}{\pi^2N_F^2}\sum_{m=-\infty}^{\infty}\sum_{m'=-\infty}^{\infty}e^{-i(m'-m)(\phi'-\phi)}\Biggl\lbrace 
\notag\\
&\left[t_1^m(u',u)^{\ast}+(-1)^{m+1}\frac{\Delta\sqrt{1-u^2}}{E}e^{-i(\phi'-\phi)}t_2^m(u',-u)^{\ast}\right]
\times
\left[t_1^{m'}(u',u) + (-1)^{m'}\frac{\Delta\sqrt{1-u'^2}}{E}t_3^{m'}(-u',u)\right] 
\notag\\
+&\left[(-1)^{m+1}t_2^m(u',-u)^{\ast}+\frac{\Delta\sqrt{1-u^2}}{E}e^{i(\phi'-\phi)}t_1^m(u',u)^{\ast}\right]
\times
\left[(-1)^{m'+1}t_2^{m'}(u',-u) + \frac{\Delta\sqrt{1-u'^2}}{E}t_4^{m'}(-u',-u)\right] 
\notag\\
+&\left[(-1)^{m}t_3^m(-u',u)^{\ast}+\frac{\Delta\sqrt{1-u^2}}{E}e^{-i(\phi'-\phi)}t_4^m(-u',-u)^{\ast}\right]
\times
\left[(-1)^{m'}t_3^{m'}(-u',u) + \frac{\Delta\sqrt{1-u'^2}}{E}t_1^{m'}(u',u)\right] 
\notag\\
+&\left[t_4^m(-u',-u)^{\ast}+(-1)^{m}\frac{\Delta\sqrt{1-u^2}}{E}e^{i(\phi'-\phi)}t_3^m(-u',u)^{\ast}\right]
\times\left[t_4^{m'}(-u',-u) + (-1)^{m'+1}\frac{\Delta\sqrt{1-u'^2}}{E}t_2^{m'}(u',-u)\right]
\Biggr\rbrace.
\label{t_sq_av_4}
\end{align}
A key feature of the scattering rate is the dependence on the azimuthal angles for the incident and 
outgoing momenta \emph{only} in the combination $(\phi'-\phi)$. This greatly simplifies the calculation 
of the transport cross sections [see Eqs.~(\ref{sigma_+})-(\ref{sigma_-})], since any other combination 
of $\phi'$ and $\phi$ gives zero contribution after integration over incident and final state momenta. 
This allows us to show the following,
\ber\label{sigma_xy_+}
\sigma^{(+)}_{xy}(E) &=& \sigma^{(+)}_{yx}(E) = 0
\,,\\
\sigma^{(\pm)}_{xz}(E) &=&\sigma^{(\pm)}_{zx}(E)
                         = \sigma^{(\pm)}_{yz}(E)
                         = \sigma^{(\pm)}_{zy}(E) = 0
\,,
\\
\sigma^{(+)}_{xx}(E) &=& \sigma^{(+)}_{yy}(E)
\,,\\
\sigma^{(-)}_{xy}(E) &=& -\sigma^{(-)}_{yx}(E)
\,.
\label{sigma_xy_-}
\eer
To carry out calculations we project out scattering rates with difference orbital angular momenta, 
$\Delta m = 0,\pm 1$, 
\ber
\frac{2\pi}{k_F^2}W_0(u',u)
&=& 
\left(\frac{m^{\ast}}{2\pi\hbar^2}\right)^2
\int\limits_{0}^{2\pi}d\phi
\int\limits_{0}^{2\pi}
\frac{d\phi'}{2\pi}
W(\hat{\vk}',\hat{\vk})
\,,\\
\frac{2\pi}{k_F^2}W_{\pm}(u',u)
&=& 
\left(\frac{m^{\ast}}{2\pi\hbar^2}\right)^2
\int\limits_{0}^{2\pi}d\phi
\int\limits_{0}^{2\pi}
\frac{d\phi'}{2\pi}
\,
e^{\pm i(\phi' - \phi)}
\,
W(\hat{\vk}',\hat{\vk})
\eer
These rates are expressed in terms of solutions to the $t$-matrix amplitudes,
\begin{align}
&W_0(u',u) = \sum_{m=-\infty}^{\infty}\Biggl\lbrace 
\notag\\
&\left[t_1^m(u',u)^{\ast}+(-1)^{m}\frac{\Delta\sqrt{1-u^2}}{E}t_2^{m+1}(u',-u)^{\ast}\right]
\times\left[t_1^{m}(u',u) + (-1)^{m}\frac{\Delta\sqrt{1-u'^2}}{E}t_3^{m}(-u',u)\right] 
\notag\\
+&\left[(-1)^{m+1}t_2^m(u',-u)^{\ast}+\frac{\Delta\sqrt{1-u^2}}{E}t_1^{m-1}(u',u)^{\ast}\right]
\times\left[(-1)^{m+1}t_2^{m}(u',-u) + \frac{\Delta\sqrt{1-u'^2}}{E}t_4^{m}(-u',-u)\right] 
\notag\\
+&\left[(-1)^{m}t_3^m(-u',u)^{\ast}+\frac{\Delta\sqrt{1-u^2}}{E}t_4^{m+1}(-u',-u)^{\ast}\right]
\times\left[(-1)^{m}t_3^{m}(-u',u) + \frac{\Delta\sqrt{1-u'^2}}{E}t_1^{m}(u',u)\right] 
\notag\\
+&\left[t_4^m(-u',-u)^{\ast}+(-1)^{m-1}\frac{\Delta\sqrt{1-u^2}}{E}t_3^{m-1}(-u',u)^{\ast}\right]
\times\left[t_4^{m}(-u',-u) + (-1)^{m+1}\frac{\Delta\sqrt{1-u'^2}}{E}t_2^{m}(u',-u)\right]
\Biggr\rbrace
\,,
\end{align}
\begin{align}
&W_{\pm}(u',u) = \sum_{m=-\infty}^{\infty}\Biggl\lbrace 
\notag\\
&\left[t_1^{m\mp 1}(u',u)^{\ast}+(-1)^{m\mp 1}\frac{\Delta\sqrt{1-u^2}}{E}t_2^{m\mp 1+1}(u',-u)^{\ast}\right]
\times\left[t_1^{m}(u',u) + (-1)^{m}\frac{\Delta\sqrt{1-u'^2}}{E}t_3^{m}(-u',u)\right] 
\notag\\
+&\left[(-1)^{m\mp 1+1}t_2^{m\mp 1}(u',-u)^{\ast}+\frac{\Delta\sqrt{1-u^2}}{E}t_1^{m\mp 1-1}(u',u)^{\ast}\right]
\times\left[(-1)^{m+1}t_2^{m}(u',-u) + \frac{\Delta\sqrt{1-u'^2}}{E}t_4^{m}(-u',-u)\right] 
\notag\\
+&\left[(-1)^{m\mp 1}t_3^{m\mp 1}(-u',u)^{\ast}+\frac{\Delta\sqrt{1-u^2}}{E}t_4^{m\mp 1+1}(-u',-u)^{\ast}\right]
\times\left[(-1)^{m}t_3^{m}(-u',u) + \frac{\Delta\sqrt{1-u'^2}}{E}t_1^{m}(u',u)\right] 
\notag\\
+&\left[t_4^{m\mp 1}(-u',-u)^{\ast}+(-1)^{m\mp 1-1}\frac{\Delta\sqrt{1-u^2}}{E}t_3^{m\mp 1-1}(-u',u)^{\ast}\right]
\times\left[t_4^{m}(-u',-u) + (-1)^{m+1}\frac{\Delta\sqrt{1-u'^2}}{E}t_2^{m}(u',-u)\right]
\Biggr\rbrace.
\end{align}
Formulae for the in-plane transport cross sections are given in terms of integrations 
over $W_{0,\pm}(u',u;E)$,
\ber
\sigma^{(+)}_{xx}(E)
&=& 
\frac{3\pi}{4k_F^2}\int_{E\geq|\Delta(\hat{\vk})|}du
\int_{E\geq|\Delta(\hat{\vk}')|}du'\frac{E^2}{\sqrt{E^2-|\Delta(\hat{\vk})|^2}
\sqrt{E^2-|\Delta(\hat{\vk}')|^2}}\Biggl\lbrace 
\notag\\
&\times&
\left(1-\frac{u^2+u'^2}{2}\right) 
\,
W_0(u',u)
-
\frac{1}{2}\sqrt{1-u^2}\sqrt{1-u'^2}
\left[W_{+}(u',u)+W_{-}(u',u)\right]\Biggr\rbrace
\,,\label{sigma_+_xx}
\\
\sigma^{(-)}_{xy}(E)
&=& 
\frac{3\pi}{4k_F^2}\int_{E\geq|\Delta(\hat{\vk})|}du
\int_{E\geq|\Delta(\hat{\vk}')|}du'\frac{E^2}{\sqrt{E^2-|\Delta(\hat{\vk})|^2}
\sqrt{E^2-|\Delta(\hat{\vk}')|^2}}\Biggl\lbrace
\notag\\
&\times&
\frac{1-2f}{4i}\sqrt{1-u^2}\sqrt{1-u'^2}
\left[W_{+}(u',u)-W_{-}(u',u)+W_{+}(u,u')-W_{-}(u,u')\right]\Biggr\rbrace
\,.
\label{sigma_-_xy}
\eer
The other elements of the tensor cross sections are obtained by the symmetry relations, Eqs.~(\ref{sigma_xy_+}) - (\ref{sigma_xy_-}).

\section{Scattering phase shifts for quasiparticle-ion potentials}\label{appendix_phase-shifts}

For the potential defined by Eq.~(\ref{soft_sphere_waw}) the scattering phase shifts for
normal-state quasiparticles are calculated from the following expressions,
\begin{equation}
\tan\delta_l=\frac{(l-\gamma_l)j_l(k_fR')-k_fR'j_{l+1}(k_fR')}{(l-\gamma_l)n_l(k_fR')-k_fR'n_{l+1}(k_fR')}
\,,\;\;\gamma_l = x'\frac{a_l}{b_l},
\label{eq-deltas-four-parameter_potential}
\end{equation}
\begin{align}
a_l &= l\left[n_{l+1}(x)j_l(x') - n_l(x')j_{l+1}(x)\right]
+ x'\left[n_{l+1}(x')j_{l+1}(x) - n_{l+1}(x)j_{l+1}(x')\right]\notag\\
&+ \frac{lp_l}{x'}\left[n_l(x)j_l(x') - n_l(x')j_l(x)\right]
+ p_l\left[n_{l+1}(x')j_l(x) - n_l(x)j_{l+1}(x')\right],\\
b_l &= x'\left[n_{l+1}(x)j_l(x') - n_l(x')j_{l+1}(x)\right]
+ p_l\left[n_l(x)j_l(x') - n_l(x')j_l(x)\right],
\label{eq-deltas-betas}
\end{align}
and $p_l = z'i_{l+1}(z)/i_l(z)$ with
$x = \beta_1k_fR$, $x' = \beta_1k_fR'$, $z = \beta_0k_fR$, $z' = \beta_0k_fR'$,
$\beta_0 = \sqrt{\nicefrac{V_0-E_f}{E_f}}$ and $\beta_1 = \sqrt{\nicefrac{E_f+V_1}{E_f}}$,
where $i_l(x)$ is the modified spherical Bessel function of the first kind.

\twocolumngrid

\end{document}